\documentclass[acmsmall]{acmart}

\usepackage{tabularx}
\usepackage{dirtytalk}
\usepackage{comment}
\usepackage{subfig}
\usepackage{caption}
\usepackage{multirow}
\usepackage{multicol}
\AtBeginDocument{%
  }

\setcopyright{acmlicensed}
\acmJournal{PACMHCI}
\acmYear{2024} \acmVolume{8} \acmNumber{CSCW1} \acmArticle{93}
\acmMonth{4}\acmDOI{10.1145/3637370}

\begin{document}

\title[Meeting Effectiveness and Inclusiveness in Remote Meetings]{Meeting Effectiveness and Inclusiveness: Large-scale Measurement, Identification of Key Features, and Prediction in Real-world Remote Meetings}

\author{Yasaman Hosseinkashi}
\affiliation{%
  \institution{Microsoft}
  \streetaddress{1 Microsoft Way}
  \city{Redmond}
  \country{USA}}
\email{YAHOSSEI@microsoft.com}

\author{Lev Tankelevitch}
\affiliation{%
  \institution{Microsoft Research}
  \streetaddress{21 Station Road}
  \city{Cambridge}
  \country{United Kingdom}}
\email{lev.tankelevitch@microsoft.com}

\author{Jamie Pool}
\affiliation{%
  \institution{Microsoft}
  \streetaddress{1 Microsoft Way}
  \city{Redmond}
  \country{USA}}
\email{Jamie.Pool@microsoft.com}

\author{Ross Cutler}
\affiliation{%
  \institution{Microsoft}
  \streetaddress{1 Microsoft Way}
  \city{Redmond}
  \country{USA}}
\email{Ross.Cutler@microsoft.com}

\author{Chinmaya Madan}
\affiliation{%
  \institution{Microsoft}
  \streetaddress{1 Microsoft Way}
  \city{Redmond}
  \country{USA}}
\email{Chinmaya.Madan@microsoft.com}

\renewcommand{\shortauthors}{Hosseinkashi et al.}
\begin{abstract}
Workplace meetings are vital to organizational collaboration, yet relatively little progress has been made toward measuring meeting effectiveness and inclusiveness at scale. The recent rise in remote and hybrid meetings represents an opportunity to do so via computer-mediated communication (CMC) systems. Here, we share the results of an effective and inclusive meetings survey embedded within a CMC system in a diverse set of companies and organizations. We correlate the survey results with objective metrics available from the CMC system to identify the generalizable attributes that characterize perceived effectiveness and inclusiveness in meetings. Additionally, we explore a predictive model of meeting effectiveness and inclusiveness based solely on objective meeting attributes. Lastly, we show challenges and discuss solutions around the subjective measurement of meeting experiences. To our knowledge, this is the largest data-driven study conducted after the pandemic peak to measure, understand, and predict effectiveness and inclusiveness in real-world meetings at an organizational scale.
\end{abstract}




\maketitle

\section{Introduction}\label{sec:intro}
Workplace meetings are vital to collaboration and coordination in organizations \cite{allen_key_2022}. Such meetings may range from recurring team stand-ups to information-sharing, planning, decision-making, and brainstorming meetings, among others \cite{standaert_empirical_2016, a_allen_understanding_2014, romano_meeting_2001}. They form a crucial way in which individuals and organizations collaborate and engage in sensemaking, ritual, and strategic change, as well as experience stress \cite{cliff_five_2015}. Meetings, and how they are run, affect both productivity and employee well-being \cite{romano_meeting_2001, rogelberg_not_2006}. Given their importance, understanding and evaluating the quality of meetings has been a focus of the small yet growing field of meeting science \cite{romano_meeting_2001,leach_perceived_2009,cohen_meeting_2011,allen_key_2022,cutler_meeting_2021,constantinides_future_2022}. Two key dimensions of workplace meetings are meeting\textit{ effectiveness}, defined here pragmatically as the attainment of business goals, and meeting\textit{ inclusiveness}, defined here as the extent to which participants feel they have an opportunity to contribute and all voices have equal weight \cite{leach_perceived_2009,cohen_meeting_2011, allen_key_2022, constantinides_future_2022,cutler_meeting_2021}. Measuring and understanding meeting effectiveness and inclusiveness, and their drivers, is the prerequisite to improving workplace meetings through better practice guidelines and interventions.

The need to improve workplace meetings has become more imperative after the swift rise in remote and hybrid work, triggered by the COVID-19 pandemic, which has both increased the number of meetings that people have and shifted more meetings to all-remote or hybrid formats \cite{bloom_hybrid_2022,microsoft_2022_2022}. This shift, enabled by video conferencing and other computer-mediated communication (CMC) systems, also provides an opportunity to understand and improve workplace meetings in a deep and scalable way using the rich data and automation afforded by CMC systems. Here, we use novel, large-scale data from real-world meetings in a leading CMC system to address the following research questions: 

\begin{itemize}
    \item {\bf RQ1: How can we accurately measure meeting effectiveness and inclusiveness with a survey at scale and in the real-world context of meetings, as they occur during the workday?} Measuring meeting effectiveness and inclusiveness enables identifying key drivers and their interaction (see RQ2 below).  Moreover, measurement that is large-scale (i.e., with sufficient statistical power) and real-world (i.e., occurring in the local context of meetings) provides metrics necessary to track changes in an organization's meeting culture, enabling individual organizations to improve their meetings by contextually understanding their own meeting practices and evaluating their own interventions. Critically, deploying surveys in organizations inevitably has implications for data quality, hence our focus here on accurate measurement.
   
    \item {\bf RQ2: What are the common drivers of meeting effectiveness and inclusiveness across organizations, as measured objectively, and how do they interact?} Although prior work has examined drivers like agenda use and meeting promptness (discussed in Section \ref{sec:relatedwork}), the role and interaction of drivers like length, meetings' recurring status, and attendee participation, remains unclear. Moreover, being conducted before this era of hybrid work, most prior research has not explored the role of relevant factors like audio and video participation, among others. Finally, prior research has not measured these drivers objectively and during real-world workplace meetings, and has not studied whether they generalize across organizations with different meeting experiences. These insights are necessary to develop interventions to improve workplace meetings.
    
    \item {\bf RQ3: Can a single statistical model be generalized to predict meeting effectiveness and inclusiveness for individual meetings, across organizations and industries, without relying on regular survey measurement?} Predicting meeting effectiveness and inclusiveness passively (i.e., without the need for regular survey data collection) can empower more organizations with effective metrics for tracking and improving meeting experiences. However, the feasibility and accuracy of model-based metrics for effectiveness and inclusiveness requires verification with large-scale training data and precise testing, which has not been done before.
\end{itemize}

At the heart of our contribution is scaling up the subjective measurement of effectiveness and inclusiveness in remote meetings\footnote{Remote meeting in this work refers to a meeting where participants join the meeting via a CMC system. Our data does not include telemetry about the location where participants join meetings (e.g., from home or a meeting room).} by using integrated survey ratings collected within a CMC system across multiple organizations, and linking these measurements with objective meeting attributes available via the same system (e.g., video usage, meeting size, meeting duration, etc.). To do this, we build on and substantially advance the previously developed multivariate graph model of factors associated with meeting effectiveness and inclusiveness introduced by Cutler et al.~\cite{cutler_meeting_2021}. The graph model introduced in \cite{cutler_meeting_2021} is a descriptive\footnote{Descriptive models refer to statistical models that are developed for the purpose of inferring relations among independent and outcome variables. On the other hand, predictive models are statistical models focused on inferring the values of outcome variables given the independent variables whenever the outcome variable is not known.} model that can identify a network of statistically meaningful correlations between survey ratings and meeting attributes. 
\cite{cutler_meeting_2021} demonstrates an initial proof-of-concept for a meeting effectiveness and inclusiveness survey that is integrated into a CMC system (in-client survey). 
The in-client data collection in \cite{cutler_meeting_2021} was conducted within a subset of one organization for a short time during the early stage of COVID-19, before the new norms of remote and hybrid work settled. While findings confirmed the usefulness of this approach to measure meeting effectiveness and inclusiveness, they are based on a limited set of telemetry and a specific target population with unknown generalizability to other organizations or time periods. Here, we significantly extend this work in several ways: 

\begin{enumerate}
    \item We implement a scalable measurement method in a CMC system that uses surveys to collect subjective ratings of meeting effectiveness and inclusiveness at the end of meetings 
    that can be scaled to collect data across organizations of any size. Using this system, we collect a real-world meeting dataset from five large organizations across a range of industries. This dataset was collected in 2022, after the peak of the COVID-19 pandemic, when the workforce had time to adapt its behavior to the new hybrid work context and related changes in collaboration norms. The final dataset contains 15K ratings from the rollout of this in-client survey as a new feature of the CMC system. 
    \item Using our large-scale dataset, we leverage and expand our descriptive graph model, including the range of considered meeting attributes, to conduct analyses of telemetry-captured meeting attributes and examine their interactions to contextualize and refine their relationship with meeting participation, effectiveness, and inclusiveness.
    \item We test whether our approach—the descriptive graph model based on integrated survey measurement and meeting telemetry—generalizes across organizations in different industries and with different sizes, teams, and therefore meeting experiences.
    \item To ultimately obviate the need for meeting rating surveys and thereby expand the scalability of our approach, we explore a predictive model that can predict meeting effectiveness and inclusiveness based solely on telemetry-captured meeting attributes, using survey ratings as ``ground truth'' training data.
    \item  Given the central role of survey measurement in our approach, we report on and address a key data quality issue with subjective measurement of meeting effectiveness and inclusiveness: rating skew (i.e., the tendency to avoid poor ratings). We share experiments and analysis results that aim to address this significant data quality challenge.
\end{enumerate}

Our descriptive modeling shows that, although the meeting rating baseline varies by organization, there exists a robust and consistent set of dependencies and priorities with a meaningful correlation with the effectiveness and inclusiveness of remote meetings across different organizations. This implies that the main factors and priorities related to meeting effectiveness and inclusiveness do not strongly depend on the industry or other organization-specific factors. Specifically, we find that the strongest predictor of meeting inclusiveness and hence effectiveness is whether the attendees vocally participate in the conversation. We also find that turning on video in small meetings (less than 8 participants) is correlated with a 6\% increase in the probability of participation. In the absence of video, using a headset corresponds to a 20\% increase in the odds of participation. We also show that it is harder to maintain inclusive and effective meetings with a large number of attendees: every 2 new participants corresponds to a 1 percentage point drop (absolute) in the meeting effectiveness and inclusiveness rating. Lastly, while call reliability is necessary for enabling participation and an inclusive environment, it needs to be accompanied by call quality to ensure both inclusive and effective meetings.

Moreover, we demonstrate that the descriptive model readily generalizes across organizations in different industries, with different sizes, and thereby with different meeting experiences. This is crucial because it tests the generalizability not only of our meeting-related insights but also of our measurement and modeling approach, suggesting that it can be deployed to support a wide range of organizations. In contrast, we find that predictive models that try to predict the experience of a specific attendee have lower performance than non-specific attendee models. We also show that transferring the predictive model that estimates individual ratings from one organization to another can come with a considerable drop in accuracy. 

Finally, given the central role of survey measurement in our approach, we show that survey rating skew poses a significant challenge in measuring meeting effectiveness and inclusiveness. This impacts the reliability and utility of metrics that companies can rely on to improve meeting culture. We demonstrate via analyses and experiments that rating skew can be induced by certain characteristics of the survey and its deployment, is common across organizations, and can be mitigated with survey design choices.

Our results shed light on key factors that should be considered when planning to improve meeting experiences and culture in large organizations that use CMC systems in the new era of hybrid work. We also demonstrate the value and feasibility of our measurement and modeling approach deployed in real-world contexts at an organizational scale. 

The rest of this paper is organized as follows: Section \ref{sec:relatedwork} reviews the related work in measuring meeting effectiveness and inclusiveness in meetings joined with a CMC system. Section \ref{sec:method} describes the design and execution of the in-client survey for five different corporations from different industries, the dataset variables, and the modeling methodology. The model developed and data-driven insights about the characteristics of effective and inclusive meetings (EIM) are discussed in Section \ref{sec:graph}. Section \ref{sec:skew} discusses the survey skew and its solutions. Lastly, we summarize our findings, discuss theoretical and design implications, and explore current limitations and future work.

\section{Related Work}\label{sec:relatedwork}
\subsection{Meeting design and meeting effectiveness}
Studying meetings has been a key focus of the small but growing field of meeting science (recent reviews are given in \cite{mroz_we_2018,allen_key_2022}). One survey-based method to understand the drivers of meeting effectiveness is to look for associations between meeting characteristics and participants' perceived meeting effectiveness for their recent or typical meetings. Leach et al.~\cite{leach_perceived_2009} show that, among other features, the use of an agenda, quality of facilities, and ending on time were correlated with perceived effectiveness, whereas meeting size and length were not (N = 958 survey). Attendee involvement in the meeting mediated the observed relationships. Cohen et al.~\cite{cohen_meeting_2011} examined `meeting quality', a construct similar to effectiveness (N=367 survey), finding that the top significant drivers are meeting space, size, starting promptness, lighting quality, and organization type. Although meeting size was a significant factor here, length was again not. Allen et al.~\cite{allen_meeting_2020} show meeting size to be negatively correlated with perceived meeting effectiveness. Likewise, Standaert et al.~\cite{standaert_empirical_2016} found meeting size and duration to be negatively correlated with perceived meeting effectiveness, though only for certain meetings (e.g., those using telepresence).

In summary, prior research has illuminated various key drivers of meeting effectiveness, though the role and interaction of meeting size and length remains unclear. More importantly, the role of participation in meetings and its interaction with other factors has not been rigorously studied \cite{allen_key_2022}. Leach et al.~\cite{leach_perceived_2009} found that attendee involvement is a mediator of effectiveness, but this was measured using a survey asking about overall participation across attendees (e.g., ``Participation is widespread among meeting attendees''), rather than measured objectively for individuals. This is important given that participation can vary widely among attendees, and is, therefore, difficult for one observer to retrospectively estimate overall participation \cite{cutler_meeting_2021}, \cite{lehmann-willenbrock_emergent_2016}. Additionally, no research has examined how the role of various meeting characteristics may differ between recurring and one-off meetings, given their typically different goals and structure \cite{niemantsverdriet_recurring_2017}. 

Moreover, almost all of the above research has relied on surveys administered outside the real-world context of meetings. That is, participants are asked to recall their last meeting hours later or their `typical' meeting experiences (e.g., \cite{cohen_meeting_2011, leach_perceived_2009}). This can introduce biases related to information recency or availability. Using surveys for all measurements can also introduce common-method bias \cite{podsakoff_common_2003}. Indeed, meeting science researchers have called for moving beyond solely survey methodology \cite{mroz_we_2018}. However, objective measurement of meeting characteristics (e.g., using telemetry or sensors) remains rare. Constantinides et al.~\cite{constantinides_comfeel_2020} showed that room pleasantness, as measured via sensor readings of light, temperature, etc., 
is correlated with meeting effectiveness. Cutler et al.~\cite{cutler_meeting_2021} analyzed meeting telemetry and survey rating data to show that meeting inclusiveness, the comfortableness of participating, audio/video quality, and using screen sharing and video were correlated with meeting effectiveness. A graph model was constructed with an area under the curve (AUC) of 0.68, which is promising but suggests there are many more factors to include to better understand meeting effectiveness. Lastly, given the difficulty of collecting detailed survey data, almost all prior research has relied on small samples which limits the statistical power of analyses. With the exception of \cite{cutler_meeting_2021}, to our knowledge, there have been no attempts to deploy such measurement at an organizational scale.

Equally importantly, the vast majority of prior research was conducted before the COVID-19 pandemic when norms shifted dramatically towards remote and hybrid work, which relies heavily on audio and video conferencing \cite{bloom_hybrid_2022,microsoft_2022_2022}. However, prior research has focused on characteristics of meeting facilities, such as lighting and space \cite{cohen_meeting_2011,constantinides_comfeel_2020}, rather than factors such as the use of headsets, and the choice of audio and video participation, which have become increasingly more relevant. We are not aware of studies that have examined the role of headsets and screen-sharing in workplace meetings, despite these being important factors affecting communication \cite{rintel_conversational_2010}. 

Unlike in-person meetings, remote and hybrid meetings enable participants to dynamically choose between audio and video participation throughout meetings. Studies showing the significance of video over audio conferencing are remarkably sparse and mixed. Veinott et al.~\cite{veinott_video_1999} showed that video helps non-native speakers better negotiate than audio-only conferencing. However, Habash \cite{habash_impact_1999} showed that video added little or no additional benefit over audio-only conferencing for group perception and satisfaction in distributed meetings. Instead of measuring a task metric or satisfaction, Daly-Jones et al.~\cite{daly-jones_advantages_1998} showed that video does improve conversational fluency and interpersonal awareness over audio-only meetings. Similarly, Tang et al.~\cite{tang_why_1992} showed that video conferencing system usage drops significantly when the video feature is removed, and that video is used to help mediate participants' interaction and convey non-verbal information. Sellen \cite{sellen_remote_1995} showed when remote participants join an audio or video conference with a conference room, in-room participants produced more interruptions and fewer formal handovers of the floor than remote participants (i.e., reflecting a more natural flow of conversation). However, video did not improve the interruption or handover rate for remote participants compared to audio-only. Standaert et al.~\cite{standaert_empirical_2016} showed that telepresence systems improved meeting effectiveness over audio and video conferencing systems, though they did not compare audio and video conferencing directly. More recently, Cutler et al.~\cite{cutler_meeting_2021} showed that video usage is correlated with meeting effectiveness and inclusiveness. Research is needed to understand how various forms of remote participation interact with meeting design characteristics. 

In summary, there is a need for research on meeting design and effectiveness that relies on objective metrics where possible, and that is large-scale, situated in the real-world context of workplace meetings, and is up-to-date for the new era of remote and hybrid work that depends on audio and video conferencing. 

\subsection{Measuring meeting effectiveness}
While there has been significant research in better \textit{understanding} meeting effectiveness, there is no common consensus on how to \textit{measure} it. For example, ~\cite{garcia_understanding_2019} measures meeting effectiveness by the percentage of agenda tasks that are completed. ~\cite{standaert_empirical_2016} provided 19 business meeting objectives and used a survey with a 5-point scale (1: Not at all effective to 5: Very effective) on how different meeting modalities achieved the business meeting objectives. ~\cite{nixon_impact_1992} measure meeting effectiveness using two items: goal attainment and decision satisfaction. ~\cite{rogelberg_not_2006} measure it for meetings in a typical week using a 6-item survey (5-point scale), including: “achieving your own work goals”, “achieving colleagues’ work goals”, and “promoting commitment to what was said and done in the meeting”, among others. ~\cite{leach_perceived_2009} measure it for meetings in a typical week with a 3-item survey (5-point scale): “achieving your own work goals”, “achieving your colleagues’ goals” and “achieving your department’s / section’s / unit’s goals.” 

~\cite{constantinides_comfeel_2020} measure meeting effectiveness (termed ``execution'' in their work) with one survey question (7-point scale): ``Did the meeting have a clear purpose and structure, and did it result in a list of actionable points?''. ~\cite{cutler_meeting_2021} used a single question for a large-scale email survey: ``How effective was the meeting at achieving the business goals?'' (1: ``Very ineffective'' to 5: ``Very effective''). The survey was also integrated into a CMC system, albeit using a star rating for response options (hovering over each star highlighted the associated description of each option).

In summary, whereas most prior research has included multiple aspects in measuring effectiveness, such as the perspectives of multiple attendees (e.g., one's own and colleagues' work goals \cite{rogelberg_not_2006}), or the presence of multiple features (e.g., a clear meeting structure and actionable points \cite{constantinides_comfeel_2020}), other work \cite{cutler_meeting_2021, standaert_empirical_2016} has focused on a single, pragmatic aspect: the achievement of business goals. With the exception of \cite{cutler_meeting_2021}, no prior research has deployed any measurement of meeting effectiveness at scale in a real-world meeting context where respondents may be busy or particularly influenced by work-related interactions. Further research is, therefore, necessary to understand the quality of meeting effectiveness ratings in such a context.

\subsection{Measuring meeting inclusiveness}
Meeting inclusiveness, the extent to which participants feel they have an opportunity to contribute and all voices have equal weight, is a key aspect that contributes to effectiveness \cite{cutler_meeting_2021}, \cite{constantinides_retrofitting_2021}.  \cite{nicol_l._davidson_trust_2013} reviews studies on how trust and member inclusion are factors that foster collaboration in teams, although not meetings specifically. There are many guides on how to have inclusive meetings, e.g., \cite{pendergrass_inclusive_2019}, though remarkably few studies that we are aware of actually measure inclusiveness.  

More broadly, inclusiveness has been extensively studied for organizations. ~\cite{ashikali_diversity_2013} used a large-scale (N=10,976) employee survey to build a structural equation model that shows how transformation leadership and diversity management correlate to an inclusive organizational culture (with inclusiveness measured using a six-question survey). This echoes the approach defined in \cite{ferdman_diversity_2014} (Chapter 1). \cite{pearce_expectations_2004} defined a three-question survey on measuring team inclusion. ~\cite{rice_improving_2020} further studied the relationships between organizational and supervisory inclusiveness, citizenship behavior, and affective commitment. In \cite{ferdman_diversity_2014} (Chapter 17), Lukensmeyer et al.~provide a list of important characteristics of truly inclusive meetings (discussed in detail in \cite{cutler_meeting_2021}, Section 3.1.2).  

~\cite{constantinides_comfeel_2020} measured \textit{psychological safety} (originally defined in \cite{edmondson_psychological_1999} as “the absence of interpersonal fear that allows people to speak up with work-relevant content”) using a survey question asking ``Did you feel listened to during the meeting, or motivated to be involved in it?'' (7-point scale). Inclusiveness is also related to ``group process losses'', a set of interaction dynamics identified in collaboration engineering research \cite{nunamaker_electronic_1991}. These include concepts such as ``evaluation apprehension'' (measured using survey items such as ``I felt apprehensive about expressing my ideas and findings to the rest of the group''), and ``domination'' (measured using survey items such as ``I felt that there was at least one person in the group who tended to participate much more than the other team members'') \cite{nunamaker_electronic_1991,mejias_interaction_2007}. The collective term ``group process losses'' captures the idea that these dynamics, i.e., a lack of inclusiveness, impair meeting effectiveness \cite{nunamaker_electronic_1991}.  

~\cite{cutler_meeting_2021} used a single question in a large-scale email survey: ``How inclusive was the meeting? In an inclusive meeting, everyone gets a chance to contribute and all voices have equal weight'' (1: ``Not at all inclusive'' to 5: ``Very inclusive''). Another survey question was integrated into a CMC system: ``Did you feel included in the meeting?'' (a star rating, with hover text, 1: ``I didn't feel included at all'' to 5: ``I felt very included''). 

The experience of inclusiveness may be related to individual factors such as gender. While there are many studies showing gender bias in speaking and interruption rates \cite{eecke_influence_2016,james_understanding_1993,leaper_meta-analytic_2007}, both \cite{cohen_meeting_2011,leach_perceived_2009} show that gender is not correlated to meeting effectiveness. ~\cite{triana_does_2012} showed women felt more included and participated more when CMC meetings were used before face-to-face meetings, compared to after. ~\cite{guo_effectiveness_2006} showed that traditional face-to-face meetings outperformed videoconferencing when accounting for team-building experience (but a dialogue-based framework in virtual teams can mitigate these differences). 

For meeting effectiveness, with the exception of \cite{cutler_meeting_2021}, no prior research has deployed any measurement of meeting inclusiveness at scale in a real-world meeting context. Further research is therefore needed to understand the quality of meeting inclusiveness ratings in such a context, particularly given its relatively more sensitive nature \cite{constantinides_retrofitting_2021}.

\subsection{Predicting meeting effectiveness and inclusiveness}
Given the challenges of collecting large-scale survey data, using passively and objectively captured metrics to predict meeting effectiveness and inclusiveness would enable meeting measurement to scale to entire organizations. This would afford organizations the ability to understand and improve their own meetings. However, very few studies to our knowledge have attempted to do this. ~\cite{zhou_role_2021,zhou_predicting_2022} quantified the text and vocal characteristics of meeting discussions and found that certain types of conversations (e.g., conflict, social support) and expressed emotions (e.g., disappointment, excitement) are predictive of meeting success (defined in the study as a combination of factors similar to meeting effectiveness and inclusiveness). ~\cite{choi_kairos_2021} analyzed body cues such as head and hand movements to predict meeting success (defined above). However, these approaches are not privacy-preserving and require participants to wear measurement technology, and therefore face challenges in scaling up to the organizational level. Moreover, they do not make use of the important meeting design characteristics that have been previously investigated. Further research is therefore needed to explore the feasibility of predicting meeting effectiveness and inclusiveness in a privacy-preserving way using objectively measured meeting design characteristics \cite{cutler_meeting_2021}.   

\subsection{Remote collaboration}
In addition to the structural aspects of meetings, which we address in this paper, there has been significant work done on remote collaboration and the non-structural aspects of collaboration. Olson \cite{olson_working_2013} provides a summary of recommendations for effective remote collaboration and best practices for remote meetings. Woolley et al.~\cite{woolley_evidence_2010} studied the collective intelligence of groups and showed that groups with more equal distribution of turn-taking and groups with more equal distributions of gender had a higher collective intelligence. Lykourentzou \cite{lykourentzou_personality_2016} studied how personalities affect crowd-sourced teams and found that teams without a surplus of leader-type personalities exhibited less conflict and their members reported higher levels of satisfaction and acceptance. Kulkarni \cite{kulkarni_talkabout_2015} studied massive online classes and found that the more geographically diverse the discussion groups, the better the performance of the students. Kiesler and Sproull \cite{kiesler_group_1992} studied electronic mail systems and showed how they increased the flow of information in organizations, and in particular reduced social contexts such as location, distance, time, organizational hierarchy, age, and gender. 

\section{Methodology}\label{sec:method}
The current work relies on subjective ratings of meeting effectiveness and inclusiveness collected via an in-client survey within a CMC system, together with meeting participation and attributes captured via telemetry (e.g., the meeting size, length, video usage, etc.). The CMC system randomly showed the survey at the end of remote meetings for the participating organizations in a pilot program. The resulting data, combined with our company's survey data, construct a valuable dataset from real meetings. 
We apply predictive and descriptive modeling strategies to answer the key research questions using this data. 

Section \ref{sec:survey} tackles RQ1 to some extent by presenting the design and implementation of a scalable survey tailored to measure the effectiveness and inclusiveness of meetings. The decisions made regarding the survey's design and deployment are crucial as they have substantial implications for data quality. We will discuss these implications in detail in Section \ref{sec:skew}, providing a detailed investigation to address all aspects of RQ1.

We provide insights into the pilot program in Section \ref{sec:orgs}, where we maintain the confidentiality of participating organizations. Detailed information about the dataset is available in Section \ref{sec:m-variables}. In Section \ref{sec:method_modeling} we elucidate the modeling techniques employed for the analysis and extraction of insights from this dataset.

\subsection{Survey Development and Implementation}\label{sec:survey}

In line with ~\cite{cutler_meeting_2021}, we define meeting \textit{effectiveness} as the extent to which business goals are attained, and meeting \textit{inclusiveness} as the extent to which participants feel they have an opportunity to contribute and all voices have equal weight. This approach aims to minimize subjectivity by avoiding multiple perspectives or asking about decision satisfaction and applies to a broad range of workplace meetings. Moreover, it is short and simple enough to be used in large-scale deployment within organizations. 

We designed a meeting effectiveness and inclusiveness survey that engaged meeting participants immediately after they left the meeting. This ensures that ratings are as proximate as possible to the actual meeting and therefore minimizes bias in recollection. However, it also means that post-meeting activities such as sharing meeting notes or actions do not impact the survey responses. Therefore, our operational definition of meeting effectiveness necessarily focuses only on activities that happened before or during the meeting. 

\begin{figure}
    \centering
    \subfloat[\centering Page 1]{{\includegraphics[width=7cm,trim={0mm 0mm 0mm 7.5mm},clip]{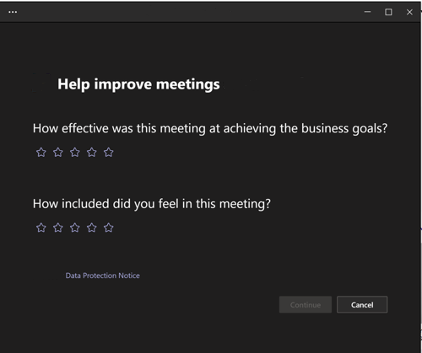} }}
    \qquad
    \subfloat[\centering Page 2]{{\includegraphics[width=7cm,trim={0mm 0mm 0mm 0mm},clip]{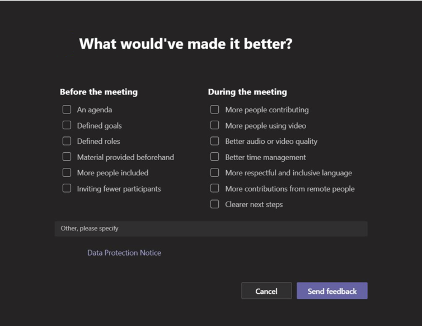} }}
     \caption{Initial End-of-Meeting Survey. We dropped the second page and only showed the first page according to randomized A/B experiment results.}
    \label{fig:survey}
\end{figure}

Initially, our end-of-meeting survey consisted of two pages as seen in Figure \ref{fig:survey}. The first page consisted of two questions:
\begin{enumerate}
    \item How effective was this meeting at achieving the business goals?
    \item How included did you feel in the meeting?
\end{enumerate}
Users could provide a rating between 1 and 5 stars, cancel out of the survey, or not provide an answer altogether, and the survey would time out after 30 seconds. The CMC system randomly selects meetings where all participants are shown the survey at the end of the call. This design provides a random sample that represents different meeting experiences in an organization. The low triggering rate of the survey reduces the chance of the same user being exposed to the survey frequently. More description about the choices of survey design is provided in section \ref{sec:experiment}.

If a user gave anything less than a 5-star rating, they would be shown a second page. The user could select one or more ``problem tokens'' from 13 options that they felt would have improved the meeting experience. Users could provide verbatim feedback in the ``Other: please specify'' text box. Like the first page, users could choose not to select anything, and the survey would time out, or users could click cancel. 

We launched this two-page survey within our organization and conducted randomized experiments with the survey design, including the interface, and survey frequency. These experiments led to changes in the design, including the elimination of the second page. The final survey that is used for the current analysis contains a single page (Figure \ref{fig:survey} (a)). 

\subsubsection{Experimentation:}\label{sec:experiment}
 Two randomized controlled experiments were designed to compare different survey interfaces and triggering logic (frequency): 

\begin{enumerate}
   \item Comparing two survey interfaces:  Control (A): The two-page survey as shown in Figure \ref{fig:survey}. Treatment (B): One-page survey (removed the second page of the survey shown in \ref{fig:survey} (b)). The control survey gathers more meeting experience details but comes with a higher cognitive load, potentially reducing data quality and quantity. Comparing both surveys helps gauge the impact of survey length on data quality in this context. This experiment was run for 6 weeks. During this time, the control and treatment populations (about 8K users combined) received the respective surveys in randomly selected $10\%$ of their meetings. The main hypothesis was that the one-page survey allows for more a diverse rating distribution (less biased and more informative data).
   \item A two-factor experiment designed to choose ``Trigger Rate'', i.e., the percentage of meetings selected for the survey, and ``cool-down period'', i.e., the minimum amount of time that is needed to lapse between two consecutive survey exposures for a user. This experiment ran for 2 weeks within a population of 4.6K users. The main hypothesis was that less frequent survey exposure leads to a more diverse rating distribution (healthier and more informative data).

The four treatments were:
   \begin{enumerate}
     \item 1\% trigger rate, no cool-down
     \item 5\% trigger rate, no cool-down
     \item 15\% trigger rate, no cool-down
     \item 15\% trigger rate, 7-day cool-down
   \end{enumerate}
 \end{enumerate}
\vskip 0.1in
The experiments were conducted in non-overlapping time periods. 
The results from the experiments led to clear conclusions that were implemented in our pilot program with external organizations:
\begin{enumerate}
\item The one-page survey enables more distributed ratings. This is demonstrated by a statistically significant drop in the responses with 5-star ratings on both questions, i.e., ``Perfect Meeting Rate'' (PMR). 
\item The 7-day cool-down also corresponds to the lowest PMR.
\item A high trigger rate, if not accompanied by a cool-down period, can lead to more skewed ratings.
\end{enumerate}
\vskip 0.1in

The one-page survey was finally implemented with a built-in 7-day cool-down period and a 10\% trigger rate. For global organizations, additional rules were implemented to follow the local rules as needed. Importantly, the relatively low trigger rate (survey frequency) and the 7-day cool-down period both minimize the impact of the survey on user behavior during meetings, as participants are not constantly asked to complete the survey.  

Telemetry collection happens entirely behind the scenes and automatically by the application without any user involvement. 

\subsection{Dataset and Pilot Program}\label{sec:orgs}
After launching the survey internally, we partnered with five global companies ranging from 5,000 employees all the way to over 500K employees to build a diverse dataset. Table \ref{tab:org_describe} gives an overview of the different types of organizations participating in this study.
\begin{table}
  
  \label{tab:freq}
  \begin{tabular}{p{0.1\textwidth}>{\raggedright\arraybackslash}p{0.2\textwidth}p{0.2\textwidth}p{0.2\textwidth}p{0.2\textwidth}}
    \toprule
     Company & Employee Count (approximate) & Industry & Countries Included& Total Responses\\
    \midrule
A  & 32,000 & Telecom &Global & 2,624 \\
 B & >500,000 & Consulting & US/Canada Only& 2,450 \\  
 C & 20,000 & Consulting & US/Canada Only& 1,306\\
 D & 20,000 & Consumer Goods& Global & 971\\
 E & 5,000 & Agriculture and \newline Construction & US Only& 66\\
  \bottomrule
\end{tabular}
\caption{\label{tab:org_describe}Description of participating organizations and dataset size.}
\end{table}
Data was collected in aggregate during the period between March 1st - June 30th, 2022, however, the exact timelines differed by participating organizations. All companies adopted hybrid policies by 2022, but the exact number of days spent in the office vs. remotely is unknown. Our data only contains ratings from users who participate in the meeting via the CMC system. Although the majority of meetings in our data are likely to be all-remote (i.e., users participate remotely via the CMC system), our data cannot identify and exclude participants that were physically co-located in hybrid meetings (see also Section \ref{sec:limitations} for discussion of this). After applying filters, approximately 7,330 usable responses from these organizations remained in the dataset. This accounts for 61\% of the unfiltered collected ratings. Filters had to be applied to ensure that ratings were valid for our analysis. The main filters are >2 participants, call duration < 150 minutes, and time taken to complete the survey > 4 seconds. The filter on call duration is in place to exclude calls that may not confirm a similar pattern, purpose, or goal as in regular meetings. This filter has removed about $1\%$ of all data points.  
The final dataset consists of both the external organization and internal pilot data: 15K ratings in total after applying filters. 

\subsection{Dataset Variables}\label{sec:m-variables}

Survey responses and call telemetry are linked using a shared identification number, protecting the privacy of both the respondent and the meeting participant. Call telemetry provides limited but secure insights into meeting attributes, including technical aspects and partial user behavior, with no personal identifiers. Consequently, demographic attributes are not included in this data. While this absence of personal information can be a challenge when modeling effectiveness and inclusiveness metrics, it also ensures that the resulting model and insights can be seamlessly integrated into real-world applications without requiring sensitive information. 

In the rest of this section, we introduce the main variables in the current work. 

\subsubsection{Outcome variables}\label{sec:outcomevars}
The survey used in the current work contains two questions about meeting effectiveness and inclusiveness on a 5-point star rating scale (Figure \ref{fig:survey} (a)). We define two binary variables from these ratings: \texttt{Effective} and \texttt{Inclusive}, defined as 1 if a 4- or 5-star rating and 0 otherwise. Every other variable used in this study comes from meeting telemetry. 
\texttt{Effective} and \texttt{Inclusive} are the only variables in this study that are provided by users and are our main outcome variables in modeling. 

We also have a telemetry-based outcome variable: \texttt{Participation}.  \texttt{Participation} is 1 if the user participates vocally in the meeting and 0 otherwise. 
It is computed based on the Number of Encoded audio Frames (NEF) throughout the call. NEF is recorded per participant and does not require any audio recording. It is merely based on counting the number of audio packets from one participant during the call. NEF is normalized by meeting size to have the same scale as the ``proportion of the meeting that the attendee spoke'' (this is because audio frames are only sent when a voice activity detector is triggered). We consider NEF > $10\%$ as a proxy for ``participating in the conversation''. So \texttt{Participation} is 1 if NEF>$10\%$ and 0 otherwise. Note that this does not mean exactly speaking for $10\%$ of the meeting duration. It is a proxy for participation that is determined based on a small user study and correlation analysis with the real data. We conducted a small user study where users provided consent to share the audio content. In these calls, we compared the duration of participating in a conversation and normalized the NEF. The data showed that less than $10\%$ of the values are the results of greetings at the beginning and end of the meeting rather than meaningful participation in the conversation. This was later confirmed by the high correlation pattern observed with other outcome metrics using the main data.

We will refer to the three variables \texttt{Effective}, \texttt{Inclusive}, and \texttt{Participation} as Effective Inclusive Meeting (\textit{EIM) outcome metrics}.

\subsubsection{Independent variables}\label{sec:attributevars}
In addition to \texttt{Participation}, telemetry provides detailed information on these areas: meeting duration, each attendee's call duration, number of participants (meeting size), choice of media\footnote{Media refers to audio, video, or screen sharing.} and its duration by each participant, minimal information on meeting's scheduling metadata such as time of day, day of the week, and type\footnote{Scheduling frequency}, whether a USB headset was in use, general statistics on network condition, and audio/video signal processing statistics. 

In our initial modeling steps, we use binary variables. To transform continuous variables into binary ones, we select thresholds or ranges that show the highest sensitivity to the EIM outcome metrics when analyzed individually: We scan across a wide range of thresholds, calculate the lift\footnote{Lift in a binary variable X by another binary variable Y is the conditional probability of X=True given Y=True divided by the overall probability of X=True. It measures how much a change in one variable is associated with a similar change in the other variable.} in the EIM outcome variables, and choose the threshold that results in the most significant lift. In the case of ties, we opt for a middle value or one that holds practical significance

For example, we measure video usage by \texttt{Video Duration Percent > 30\%}.  Since each participant in a group call can independently choose to enable or disable their video, we measure video usage on a per-participant basis. In our analysis, video duration refers to the portion of the call when a participant both viewed others' videos and shared their own. Our initial evaluations showed that video duration is only beneficial if relative to the call duration. Hence, we define video usage as a percentage of the call duration. \texttt{Video Duration Percent} varies between 0\% and 100\%. To create a binary variable, we compute the lift in \texttt{Effective} and \texttt{Inclusive} by 
\texttt{Video Duration Percent > t} for multiple values of $t\in[0.01, 0.9]$. The lift is highest when $t\approx 0.3$.  
Similarly, we found interesting binary variables by converting \texttt{Call Duration} to \texttt{Short call (10 min. or less)} and \texttt{Meeting Size} to \texttt{Small Meeting (8 or less)}.

Meeting telemetry includes the type of the meeting: recurring (repeated regularly, such as weekly), scheduled (one-off meeting invitations), or ad-hoc (calls that people in a group chat initiate without a
prior calendar invite). This study excludes ad-hoc meetings since they do not receive the Effective Inclusive Meeting (EIM) survey. We used the binary variable \texttt{Recurring} in modeling and insights development.

We also utilized the rich telemetry about network and signal processing statistics to generate composite metrics for call quality and call reliability. 

\texttt{Quality Issues} is a binary classifier that consumes 40 telemetry statistics about issues such as echo, noise, or speech distortions. We trained this classifier independently on ground truth from the Call Quality Feedback (CQF) survey that is displayed at the end of a random subset of calls in the CMC system\footnote{This survey uses a Likert scale to measure the quality of the call}. Call quality ratings are the most accurate measures of call quality available from real calls. It is a single rating that reflects the user's opinion about the overall quality and is collected immediately after the call ends.
\texttt{Quality Issues} is a gradient-boosting decision tree (lightGBM \cite{ke_lightgbm_2017}) that is trained to predict the probability of poor CQF rating (rating 1 or 2 out of) 5-stars).
We measure the performance of binary classifiers by the Area Under the Curve (AUC) of the Receiver Operating Characteristic curve. This classifier has $74\%$ AUC on the validation set. A $74\%$ AUC is considered good performance when using call telemetry to predict real user ratings on entire call quality in this CMC system.  Call telemetry is aggregated statistics and does not fully capture the changes in quality during the call. For this and other reasons, predicting user ratings just based on aggregated statistics remains a challenging task. 

\texttt{Reliability Issues} is a simple aggregation of telemetry about call drop, one-way audio, or similar problems. This metric does not require advanced machine learning solutions since most reliability problems are well detectable by the application itself, and there is already binary telemetry to record their presence. The metric \texttt{Reliability Issues} is 1 if the call involves any reliability problem, 0 otherwise. 

From a user perspective, \texttt{Reliability Issues} captures whether users can join and stay in a meeting remotely. In contrast, \texttt{Quality Issues} capture the technical audio/video \textit{quality} of their experience if they can join and stay in the meeting.                              
\subsection{Modeling Methodology}\label{sec:method_modeling}
Our main analysis applies three modeling techniques: 
\begin{itemize}
    \item EIM graphical model \cite{cutler_meeting_2021} to detect the most important attribute of \texttt{Effective} and \texttt{Inclusive} meetings and their correlation structure
    \item Generalized Linear Models (GLM) \cite{dobson_introduction_2002} to explore interaction effects in sub-graphs
    \item Gradient-Boosting Decision Tree, such as lightGBM \cite{ke_lightgbm_2017}, to test the predictive power of available telemetry for EIM metrics 
\end{itemize}  

\begin{figure}
 \centering \includegraphics[width=0.75\linewidth]{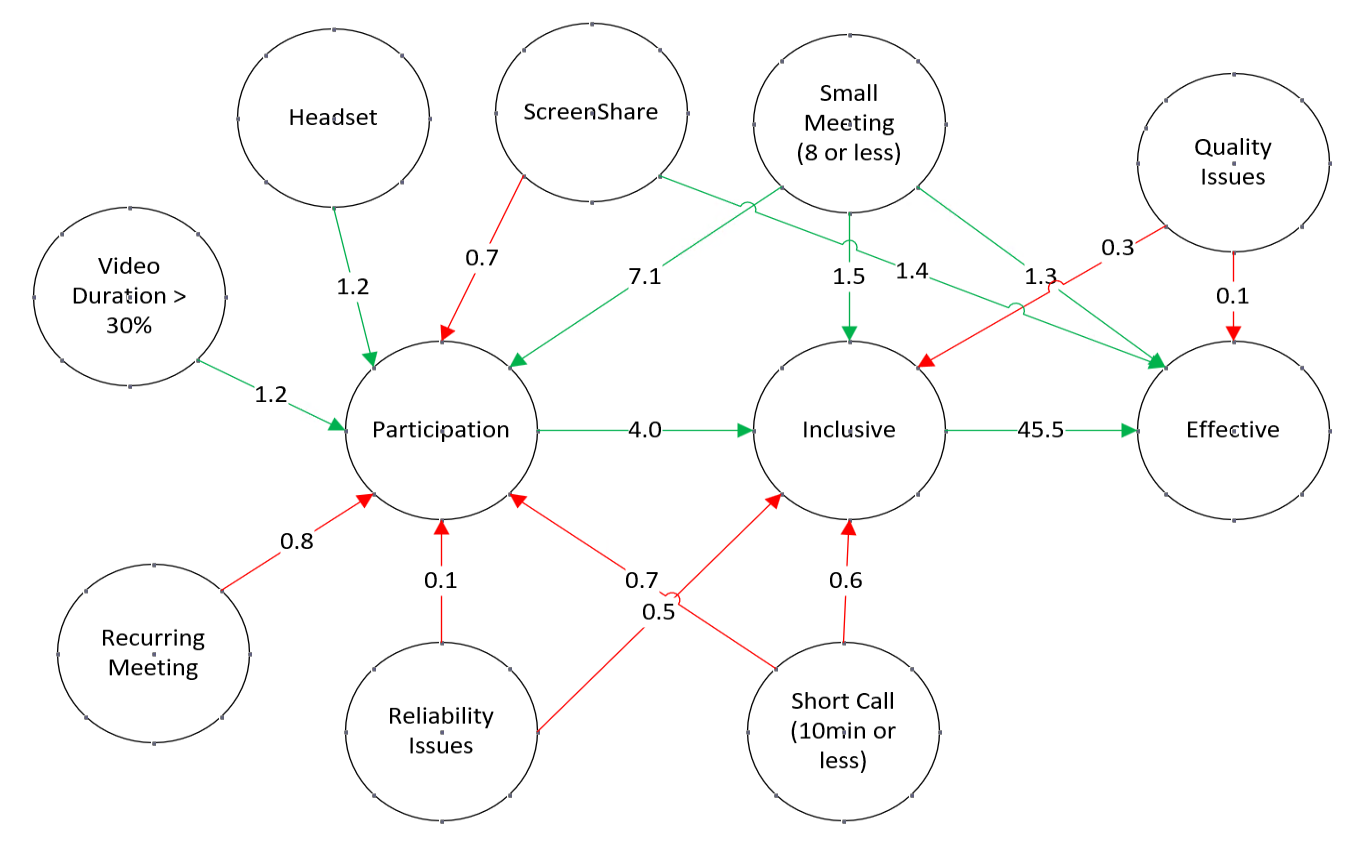}
 \caption{The graphical model showing the network of conditional dependencies between meeting attributes and \texttt{Effective} and \texttt{Inclusive}.  The red and
green edges show negative and positive dependencies, respectively. The weight on each edge is the adjusted Odds Ratio as a comparable measure of strength for each dependency.}  
 \label{fig:gm}
\end{figure}
We apply the algorithm introduced in \cite{cutler_meeting_2021} (with minor modifications) to the three EIM outcome metrics: \texttt{Participation} $\rightarrow$ \texttt{Inclusive} $\rightarrow$ \texttt{Effective}. In this graph, each node is a variable, and each directed edge represents an adjusted Odds Ratio (OR) from a multivariate GLM model. Directed edges are from independent to outcome variables. 

The algorithm to fit this model \cite{cutler_meeting_2021} has two main steps. First, the neighborhood for each node is determined using $l1-$regularized logistic regression. This is shown to provide a close approximation of an optimum graph structure (see section 3.3.5.1 in \cite{cutler_meeting_2021}). The result of this step is a graph structure or sets of neighborhoods for each main outcome variable. The algorithm is simplified by fixing the main outcome variables to \texttt{Inclusive}, \texttt{Effective}, and \texttt{Participation}.

The hierarchy from \texttt{Participation} to \texttt{Inclusive} and then \texttt{Effective} is determined by the clues from the data and literature. We set the order between \texttt{Effective} and \texttt{Inclusive} based on the Akaike Information Criteria (AIC) \cite{dobson_introduction_2002}: model AIC value is lower when \texttt{Inclusive} predicts \texttt{Effective}. A similar result is reported in \cite{cutler_meeting_2021}. \texttt{Participation} is the third outcome node in this graph because of its significant role in connecting EIM metrics with many meeting attributes. \texttt{Participation} is not only the strongest predictor of \texttt{Inclusive}, but it also correlates with more actionable attributes like video and headset usage. Without \texttt{Participation} as a link, we couldn't learn the detailed effects and interactions among these attributes. In addition to the support from data, this hierarchy aligns with prior research showing that participation is important for inclusiveness \cite{allen_key_2022} and that inclusiveness drives meeting effectiveness \cite{constantinides_comfeel_2020, nunamaker_electronic_1991}.

The second step in fitting the graph model from \cite{cutler_meeting_2021} is estimating the weights of graph edges. The coefficients of $l1-$regularized logistic regression are not proper candidates for this purpose. While the regularization technique used in this step is an effective way of reducing the number of variables, it does not provide a valid ground for statistical hypothesis testing on these parameters \cite{hastie_elements_2001}. Hence the second step of the algorithm applies GLM modeling without regularization to estimate the weights for each edge and further prune the graph if the parameters are statistically zero at a 95\% confidence level.

In a separate modeling task in this work, we use LightGBM to fit a predictive model. LightGBM is a gradient-boosting decision tree and is less interpretable than linear models. This property makes LightGBM more appropriate and successful in predictive tasks than descriptive modeling. The set of variables that are used as input for this model is larger and includes more granular variables since there are no interpretability requirements for the predictive model.

\section{Effective Inclusive Meeting (EIM) Model Results}\label{sec:graph}
This section addresses RQ2 and RQ3 by constructing models based on the connections between meeting characteristics and \texttt{Effective} and \texttt{Inclusive} metrics as described in Section \ref{sec:outcomevars}. To answer RQ2, we fit the EIM graph model using the two-step algorithm described in Section \ref{sec:method_modeling}. Initially, we considered all available binary variables, allowing the algorithm to determine their relevance as predictors in the EIM graph model. Figure \ref{fig:gm} shows the results from analyzing the entire dataset comprising 7,330 data points. Red and green edges in the figure denote negative and positive dependencies, respectively, with edge weights representing adjusted ORs computed from GLM coefficients for each node. he findings from this model and related follow-up GLM models are discussed in Sections \ref{sec:whyOdds}, \ref{sec:keyatt}, and \ref{sec:insights}.
In addition, Section \ref{sec:gen_predic} utilizes Gradient-Boosting Decision Tree modeling to address RQ3. 

\subsection{Odds Ratio as Graph Weights}\label{sec:whyOdds}

The \textbf{odds} of an event is the ratio between the probability of an event occurring and not occurring. For example, if the overall probability of a meeting being \texttt{Effective} is 90\%, then the odds of it being \texttt{Effective} is $0.9/0.1$ or ``9 to 1''. 

ORs are ratios between the odds of an event under two different conditions formed by another, e.g., the rate between the odds of \texttt{Effective} meetings with and without \texttt{Quality Issues} in the call. If there is no dependency, then the OR is close to 1. OR values greater than 1 represent positive dependency and lower than 1 represent negative dependency. ORs provide a standardized measure to compare the strength or importance of attributes regardless of any assumption about the conditional rate of the outcome variable. 

For example, 0.1 on the edge between \texttt{Quality Issues} and \texttt{Effective} means that the odds of an \texttt{Effective} meeting experience are 90\% lower when the call has quality issues. Likewise, the 1.3 between \texttt{Small Meeting (8 or less)} and \texttt{Effective} means that attendees in meetings with at most 8 participants have 30\% higher odds of having an \texttt{Effective} meeting experience. Critically, the 0.1 and 1.3 numbers are meaningful and comparable regardless of any baseline distribution for \texttt{Effective} and can be fairly compared with each other, hence our choice. However, if converting odds into probabilities, the results are only meaningful under an assumed rate of \texttt{Effective} under no \texttt{Quality Issues} and no \texttt{Small Meeting}. For example, to convert the OR=0.1 between \texttt{Quality Issues} and \texttt{Effective} into the \% change in ``the probability of \texttt{Effective}'', we need to specify the probability of \texttt{Effective} in a call without quality issues (baseline). Assuming that this baseline is 95\%, OR=0.1 means the 95\% chance of \texttt{Effective} drops by 66\% in the presence of quality issues. Note that the 66\% now is relative to the 95\% baseline and would change with a different baseline. Since the baseline is specific for each attribute, the \% change in probability of the model attributes is not necessarily comparable. Therefore, OR, being free of such assumptions, is the appropriate metric for interpreting multivariate models with binary variables. 

\subsection{Key Attributes}\label{sec:keyatt}
\texttt{Inclusive} and \texttt{Effective} have the highest correlation in the EIM graph. We believe this strong correlation is exaggerated due to common-method variance \cite{podsakoff_common_2003} and the survey response characteristics; see Section \ref{sec:skew} for an in-depth discussion of this. The weight is large enough to encourage collapsing the two variables and creating a composite outcome variable for the model. However, our experiments showed that combining them into a single outcome variable weakens the descriptive power of the model. For example, the effects of \texttt{Participation} and other predictors are statistically strongest if using \texttt{Inclusive} as a stand-alone outcome metric. Similarly, a model that predicts \texttt{Inclusive} allows for a better understanding of interactions between inputs than a composite outcome variable. Therefore, we kept \texttt{Inclusive} and \texttt{Effective} as separate nodes in the graph. 

After the \texttt{Inclusive} – \texttt{Effective} edge, the highest correlated pairs are:
\begin{enumerate}
    \item \texttt{Participation} – \texttt{Inclusive} (OR = 4.0)
    \item \texttt{Small Meeting} – \texttt{Participation} (OR = 7.1)
    \item \texttt{Quality Issues} – \texttt{Effective}/\texttt{Inclusive} (ORs = 0.1 and 0.3)
    \item \texttt{Reliability Issues} – \texttt{Participation} (OR = 0.1)
\end{enumerate}

Section \ref{sec:insights} describes a deep analysis of these areas. It is worth mentioning some of the variables that were expected to be correlated with meeting effectiveness and inclusiveness but were dropped by the algorithm. These are great examples of the importance of large data to help validate prior theories or hypotheses about what matters in meeting inclusiveness and effectiveness. Below is a list of variables whose correlation with inclusiveness and effectiveness was not strong enough to stay in the final model.

\begin{itemize}
    \item \texttt{Day of week}: Prior research shows that more multitasking during meetings, associated with less effective meetings, happens Mondays to Thursdays (compared to Fridays), and that Mondays are associated with high boredom levels at work \cite{cao_large_2021, mark_bored_2014}. Our univariate analysis indeed showed slightly higher \texttt{Effective} and \texttt{Inclusive} rates on Fridays compared to Mondays. However, the effect was not strong enough to stay in the model in the presence of more dominant factors and vanished quickly in the modeling process.
    \item \texttt{Time of day}: Prior research shows that more multitasking during meetings happens in the mornings (compared to afternoons) and that people engage in more focused work in the afternoon \cite{cao_large_2021, mark_bored_2014}. In our data, whether the meeting occurred in the morning or the evening proved to be irrelevant to the EIM outcome metrics. The flexible work hours during the pandemic may have caused this pattern. 
    \item \texttt{Busy day}: Prior research shows that people multitask during meetings to catch up on their workload (including a high number of meetings) \cite{cao_large_2021}. In our data, the rating distributions are not different for people who have a large number of meetings (10 or more calls) on that day vs. people with fewer meetings per day. 
\end{itemize}

\subsection{Insights}\label{sec:insights}
\subsubsection{Participation:} 
\begin{figure}[h]
 \centering
 \includegraphics[width=0.6\linewidth]{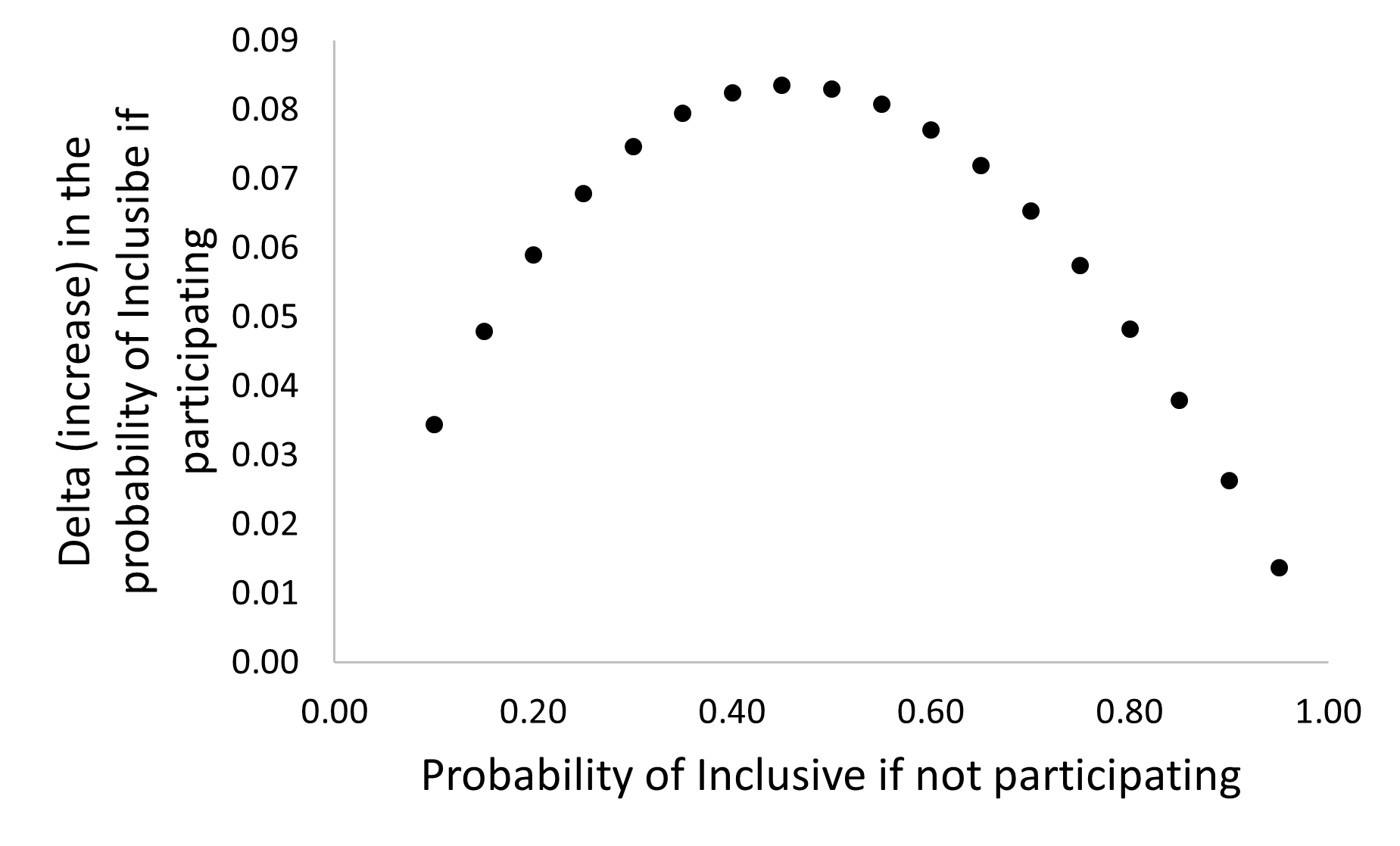}
 \caption{Model predictions for the impact of \texttt{Participation} in the probability of \texttt{Inclusive} experience for different \texttt{Inclusive} rate baselines. The impact declines for participants who would have had a very high \texttt{Inclusive} rate (over $60\%$, already very good experience) or a very low \texttt{Inclusive} rate (lower than $35\%$, significantly poor experience) without participating in conversations.}  
 \label{fig:participation_inclusibve_prob}
\end{figure}

As shown in Figure \ref{fig:gm}, \texttt{Participation} is the most important predictor of the \texttt{Inclusive} and subsequently \texttt{Effective} nodes in this model. This is in line with qualitative and survey research showing that meeting participation is key for meeting satisfaction and overall employee engagement (reviewed in \cite{allen_key_2022}). Participating in conversation is associated with 4x higher odds of having an \texttt{Inclusive} experience. This is equivalent to an 8\% higher probability of having an \texttt{Inclusive} experience for attendees whose rate of having an \texttt{Inclusive} experience is 35\% to 60\% when not participating. Interestingly, this delta is smaller for attendees with baselines below 35\% or above 60\%.  Figure \ref{fig:participation_inclusibve_prob} shows the details of this change. The figure displays a lower impact of participation when the baseline for inclusiveness is not already very high or very low. When the \texttt{Inclusive} rate is already quite high, it is not surprising that participation has limited scope for impact. However, when the \texttt{Inclusive} rate is very low, this suggests that factors other than \texttt{Participation} may be more important (e.g., the tone of meetings or other aspects of team culture \cite{allen_key_2022}). It is crucial for organizations to know their baseline \texttt{Inclusive} rate before starting any campaign for improving meeting culture. 

\textbf{Insight:} Vocally participating in meetings is associated with the largest change in the probability of having an inclusive experience (8\% increase) for meetings with a mid-range baseline probability of being \texttt{Inclusive} (35-60\%).

The strong link between \texttt{Participation} and \texttt{Inclusive} (and thereby \texttt{Effective}) motivated us to analyze the extent to which \textit{all} attendees participated in a meeting (rather than an individual) and how this correlates with attendees' experiences (see Section \ref{sec:participation}). 

\subsubsection{Meeting Size:} The second strongest correlation belongs to \texttt{Meeting Size}. The model predicts that attendees in meetings with less than eight people have respectively 50\% and 30\% higher odds of rating meetings as \texttt{Inclusive} and \texttt{Effective}. Also, they are more likely to participate by a large margin (7 times higher odds). 

\begin{figure}[h]
 \centering
 \includegraphics[width=0.8\linewidth]{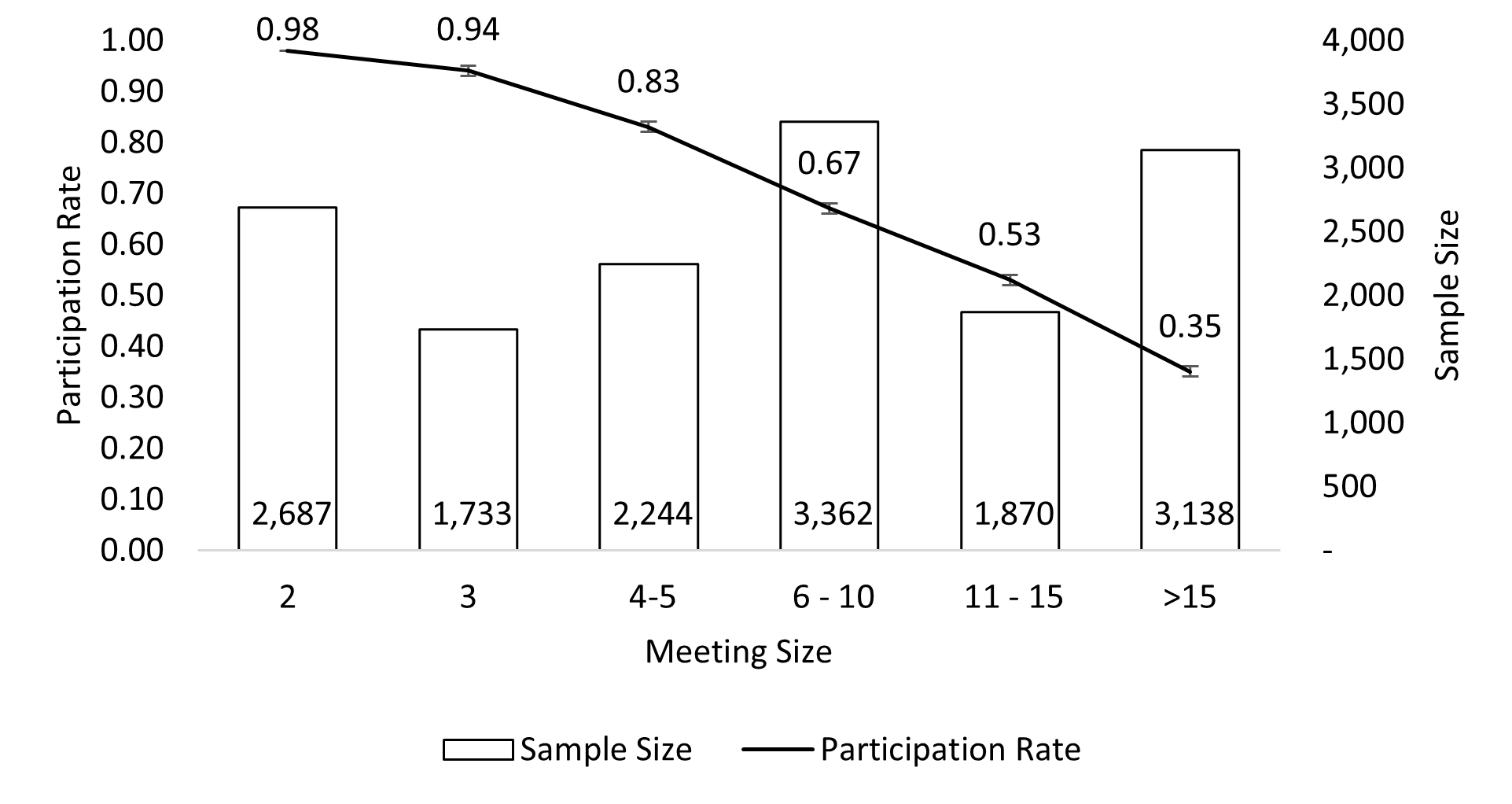}
 \caption{\texttt{Participation} rate drops as \texttt{Meeting Size} increases. The rapid decline begins with more than 5 people in the meeting. The difference in \texttt{Participation} is largest and statistically significant when we compare 8-or-less-participant meetings with more-than-8-participant meetings.}  
 \label{fig:participation_by_meetingsize}
\end{figure}

Our findings extend earlier qualitative and survey research on meeting effectiveness \cite{cohen_meeting_2011,geimer_meetings_2015, allen_meeting_2020}. Large meetings may reflect situations in which some invited participants are not relevant to the meeting, leading to real or perceived inefficiencies in time use among some or all participants \cite{cohen_meeting_2011,geimer_meetings_2015, romano_meeting_2001}. Larger meetings also provide fewer opportunities to participate (see Figure \ref{fig:participation_by_meetingsize}). This is in harmony with the findings in \cite{romano_meeting_2001}. Given the aforementioned link between participation and inclusiveness, it is not surprising that larger meetings are also associated with lower perceived inclusiveness.

We sought to better understand the link between \texttt{Meeting Size} and \texttt{Effective} and \texttt{Inclusive} meetings under different scenarios. Specifically, we were interested in the linear effect of meeting size (i.e., rather than a binary variable defining small meetings as 8 attendees or less) to gain a more granular understanding. We were also interested in how this varied by whether a meeting was part of a recurring series or whether it was a one-off meeting, as these are broadly associated with different meeting purposes (e.g., team stand-ups vs. brainstorming meetings) that may be differently impacted by meeting size \cite{allen_key_2022}. Additionally, we looked at the role of meeting duration, as prior research shows that long meetings are perceived as less effective \cite{allen_key_2022}. To do these analyses, we fit separate GLM models to predict \texttt{Effective} and \texttt{Inclusive} meetings using \texttt{Meeting Size} in its numeric format (not binary), with \texttt{Recurring} and \texttt{Call Duration} as inputs. Given that the above aspects can all interact in meaningful ways (e.g., a long and large recurring meeting may have a lower effectiveness than a long and large one-off meeting), we include interaction terms in the model. GLM parameters for the \texttt{Effective} outcome variable are available in Table \ref{tab:glm_par_eff} in the appendix. The results demonstrate lower \texttt{Effective} and \texttt{Inclusive} rates for larger meetings. GLM predictions under different scenarios show that if we increase the \texttt{Meeting Size} from 2 to 14, then the \texttt{Inclusive} rate falls from about 98\% to 94\% and, the \texttt{Effective} rate falls from 97\% to 93\%. However, this decline is not identical for all types of meetings. The model estimates the largest negative delta for recurring meetings that take no longer than 30 minutes; for these meetings, every two new participants are associated with a reduction in the chance of an \texttt{Inclusive} or \texttt{Effective} experience by 1\% absolute.

\textbf{Insight:}  Meetings with fewer attendees are associated with much higher odds of vocal participation and of being rated as \texttt{Effective} and \texttt{Inclusive}. The impact of meeting size on effectiveness and inclusiveness is strongest for short recurring meetings (up to 30 minutes). 

One prominent example of a short, recurring meeting is the daily stand-up meeting. Small-scale studies of daily stand-ups in software teams suggest that \texttt{Meeting Size} is indeed an important factor in the perceived success of such meetings \cite{stray_daily_2016,stray_obstacles_2013}. Here we corroborate and generalize these findings to a range of teams and industries, and identify specific quantitative effects of meeting size that are actionable.   

\subsubsection{Call Quality and Reliability:} The third group of dominant factors in the model is the quality and reliability of the call (the \texttt{Quality Issues} and \texttt{Reliability Issues} nodes).  Reliability issues occur twice more often than quality issues in this data, with strong connections to the \texttt{Participation} and \texttt{Inclusive} nodes. Given that reliability refers to basic task completion (merely being able to participate in conversation), we expect it to be necessary for a good meeting experience. But is it sufficient? The EIM graphical model demonstrates an interesting order between \texttt{Quality Issues} and \texttt{Reliability Issues}. To better reveal this pattern, we show the relevant graph weights in Table \ref{tab:QualReli}, with rows and columns indicating connecting nodes. It shows that as we move from the basic task completion (\texttt{Participation}) to ultimately \texttt{Effective} and \texttt{Inclusive} meetings, we move from a high \texttt{Reliability} correlation to a balanced mix with \texttt{Quality} and finally to just \texttt{Quality} as a highly relevant factor. 

\textbf{Insight:} Whereas call reliability is critical for enabling participation and an inclusive meeting experience, call quality becomes essential to achieve both inclusive and effective meetings.

\begin{table}[h]
\begin{tabularx}{0.75\textwidth}{llll}
\hline
 & \texttt{Participation} & \texttt{Inclusive} & \texttt{Effective}\\
\hline
\texttt{Quality Issues} & - & 0.35 & 0.14\\
\hline
\texttt{Reliability Issues} & 0.13 & 0.49 & -\\
\hline
\end{tabularx}
\caption{\label{tab:QualReli}Pattern of \texttt{Quality} and \texttt{Reliability} weights for connecting nodes in the graphical model. Rows and columns indicate connecting nodes. The empty cells (-) show that there is no edge between the corresponding two nodes in the graph.}
\end{table}

\begin{figure}
 \centering
 \includegraphics[width=0.9\linewidth]{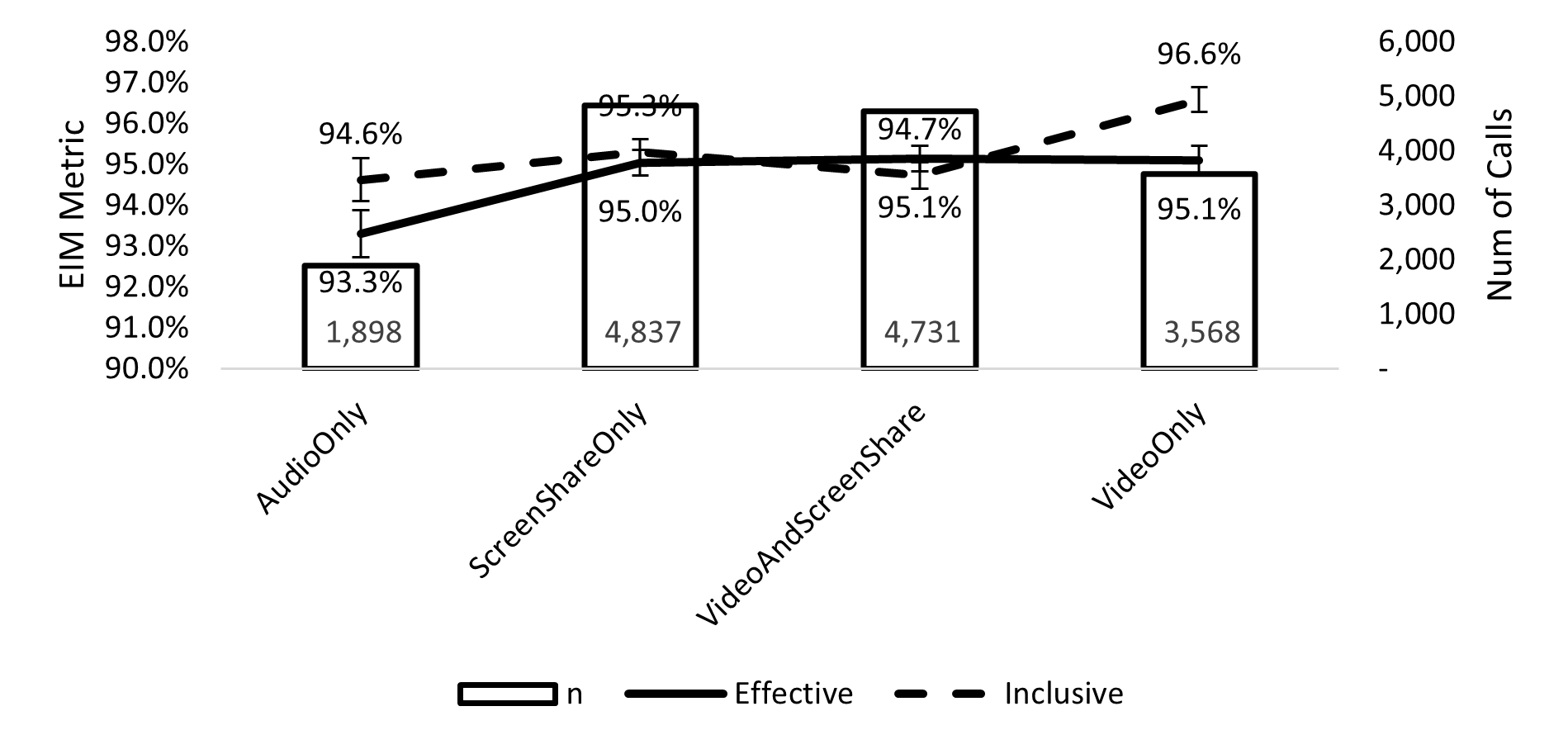}
 \caption{EIM metrics by media. The least \texttt{Effective} and \texttt{Inclusive} experiences are audio-only calls. Video-only calls are most \texttt{Inclusive} ($2\%$ higher than audio-only), but not the most \texttt{Effective}. $95\%$ Confidence intervals are shown for each metric to enable comparisons. }  
 \label{fig:media}
\end{figure}

\subsubsection{Media:} 
A remote participant has different ways of participating in meetings (i.e., different media choices):
\begin{enumerate}
    \item Audio-only: only speaking via the microphone
    \item Video: turning on their webcam video or seeing other attendees' videos
    \item Screen-sharing: sharing their screen or seeing other attendees' screens
\end{enumerate}
About 87\% of calls in this data involve participants using video or screen-sharing for various portions of the call duration. How does meeting inclusiveness and effectiveness vary with different media choices? Figure \ref{fig:media} shows effective and inclusive ratings for different media choices. We find that audio-only calls are the least \texttt{Inclusive} and the least \texttt{Effective}, while the most \texttt{Inclusive} calls are video calls, where such calls are 2\% (absolute) more \texttt{Inclusive} than audio-only calls.

Comparing video and audio-only calls shows that the \texttt{Participation} rate grows when video duration increases. To analyze this further, we look at the proportion of the call duration that involved video, defining \texttt{Video Duration Percent} as the rate of video duration divided by call duration. For example, if a participant joins a meeting for 30 minutes and uses video for 15 minutes, we see a 50\% \texttt{Video Duration Percent}. We quantify the effect of \texttt{Video Duration Percent} on \texttt{Participation} by the lift in the probability of \texttt{Participation}. We find that for an average meeting, as video duration changes from less than 10\% of the call to greater than 70\%, the lift in the probability of participation increases from ~8\% to 11\%. In other words, longer video duration is associated with a larger increase in the probability of participation.
   
While this data demonstrates video as one of the attributes of inclusive meetings, it also indicates that not all calls may benefit from this feature. The lift in \texttt{Participation} depends on the \texttt{Meeting Size} and \texttt{Call Duration}. To investigate this, we used a GLM model that predicts \texttt{Participation} probability by \texttt{Meeting Size} and \texttt{Call Duration}. We determined the thresholds for meeting size (8 attendees) and video duration (30 minutes) by repeating a similar sensitivity analysis that we used to generate binary variables out of continuous ones for the graph models. Table \ref{tab:glm_vid_par} in the Appendix contains the details of the GLM model. We find that using video is associated with the largest lift in \texttt{Participation} rate in short meetings with few participants: in a meeting with less than eight participants that takes 30 minutes or less, the chance of participation can increase by more than 6\% if attendees use video for at least 30\% of call duration. For larger meetings, the associated change in participation can even become negative when the meeting duration passes ~40 minutes.

\texttt{ScreenShare} is the only attribute with positive and negative correlations in the EIM graph (with \texttt{Effective} and \texttt{Participation} nodes). \texttt{ScreenShare} is a feature that allows a participant to present her screen during the meeting or see other people's screens. Meetings with screen-sharing have 30\% lower odds of \texttt{Participation} while 40\% higher odds of \texttt{Effective} rating. \texttt{Meeting Size} is partly explaining this pattern: meetings that have at least 10\% of their time on screen-sharing are 50\% larger than meetings with less or no screen-sharing on average. Additionally, during meetings with screen-sharing, the presenter generally talks more than the audience, resulting in lower overall participation across meeting participants. 

\textbf{Insight:} Longer video duration by meeting participants is associated with a larger increase in the probability of participation. Using video in meetings is most strongly associated with an increase in the probability of participation in short meetings (30 minutes or less) with few participants (less than eight participants). Meetings with screen-sharing are associated with higher odds of being rated as effective, and with a lower rate of participation (at least partly due to these meetings tending to be larger in size, and involving presentations in which presenters speak more than other participants). 

Our results support the notion that video plays a significant role in establishing a social presence, the ``sense of being with another'' \cite{biocca_toward_2003}, which is associated with multiple positive communication outcomes such as trust and enjoyment \cite{oh_systematic_2018}. Other studies also report that video is primarily effective in smaller groups where it can afford a sense of intimacy among participants \cite{balogova_how_2022,baym_collaboration_2021}. 

\subsubsection{Headset:} According to the graph model, the odds of \texttt{Participation} increase by ~20\% if the user is using a headset. Further analysis shows that a specific group of calls drives this correlation: audio-only or screen share-only calls, i.e., calls without video. Audio-only calls with headsets have about 5\% higher \texttt{Participation} rates than audio-only calls without a headset. 

It is worth mentioning that detecting headset usage from telemetry is a challenging task because many devices report incorrect telemetry. Therefore, we expect our “no-headset” label to have contaminated some headset users. For this reason, the current predictions may underestimate the effect of the headset on participation in no-video meetings.     

\subsubsection{Per-Meeting Participation Rate}\label{sec:participation}

\begin{figure}
 \centering
 \includegraphics[width=0.8\linewidth]{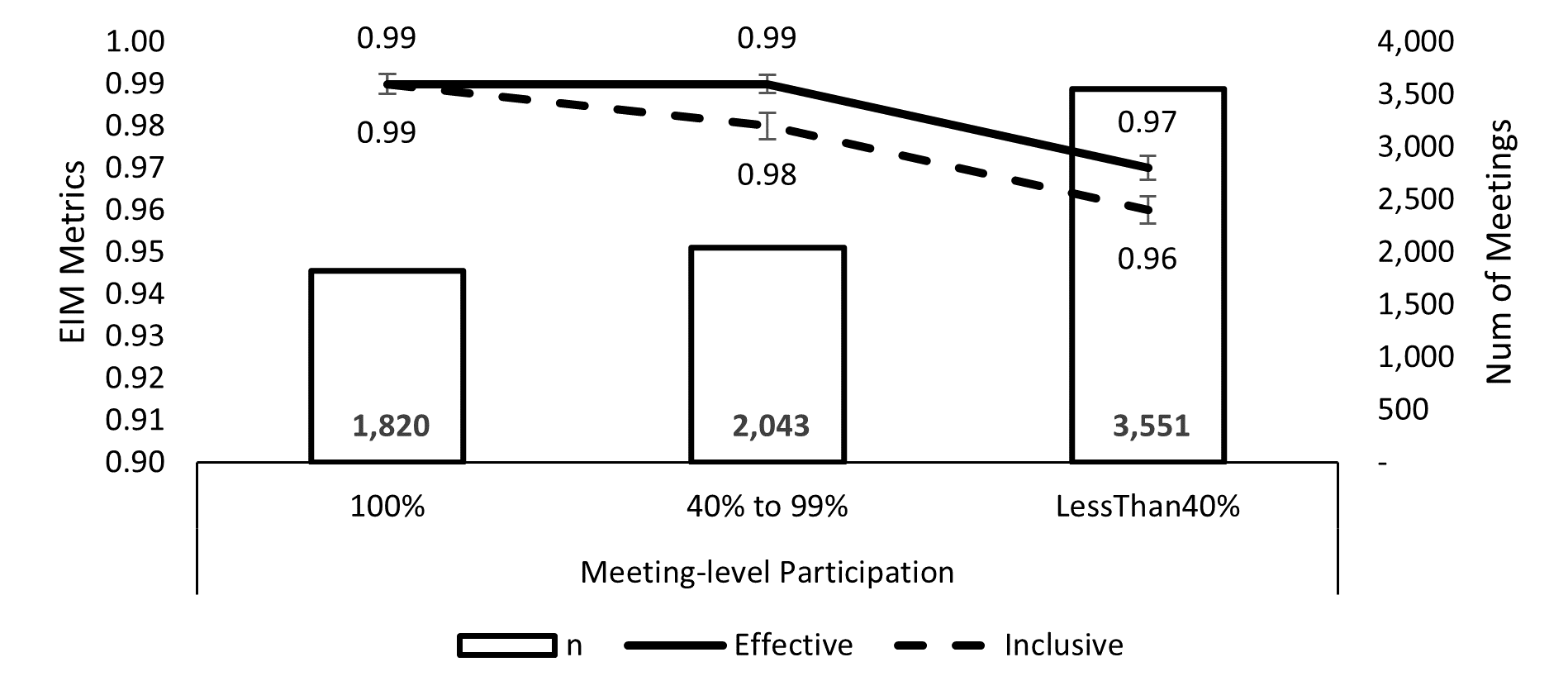}
 \caption{EIM metrics by \texttt{Participation} rate per meeting. Meeting with less than $40\%$ participation are up to $2\%$ less \texttt{Inclusive} and \texttt{Effective}}
 \label{fig:par_meeting}
\end{figure}

All previous analyses on \texttt{Participation} and its effect on the rate of \texttt{Inclusive} meetings are about correlations between attendees' ratings and their call telemetry. It shows that if one person participates in the conversations, there is a higher chance that they rate meetings as \texttt{Inclusive}. However, previous research suggests that the choice to participate in meeting conversations may be influenced by social norms set by other participants \cite{baym_collaboration_2021}. We investigated this possibility and looked at the impact of \textit{every attendee} participating on effectiveness and inclusiveness ratings. This requires aggregating both \texttt{Participation} and survey ratings for each meeting.

Participation data is available for all meeting attendees, but ratings are usually provided by at most one of the meeting attendees due to the participants' non-response\footnote{Non-response refers to missing data in surveys where respondents decide not to fill out or submit their answers. In the current research, everyone in the meeting receives the survey, so all attendees with no ratings are considered a non-response.}: only 3\% of the meetings have more than one rating. Therefore, in our dataset, meeting-level ratings are not very different from user-level ratings. However, whenever there is more than one rating available for a meeting, we use their average as the meeting-level rating of effectiveness and inclusiveness. On the other hand, we use three bins to aggregate \texttt{Participation} at the meeting level. These are the three bins for per-meeting \texttt{Participation} metric: 
\begin{itemize}
    \item 100\% (all attendees participated): a proxy for ``everyone spoke more than once''
    \item 40\% - 99\% of attendees participated
    \item Less than 40\% of attendees participated
\end{itemize}   
Figure \ref{fig:par_meeting} shows how both \texttt{Effective} and \texttt{Inclusive} rates decline together with the \texttt{Participation} rate across the three bins. We find that meetings where everyone participates are 1.8\% more \texttt{Effective} and 2.6\% more \texttt{Inclusive} than meetings where everyone doesn't participate. By contrast, meetings with less than 50\% participation are 1.5\% less \texttt{Effective} and 2.7\% less \texttt{Inclusive} than meetings with more or equal to 50\% participation.

\textbf{Insight:} The extent of participation \textit{across participants within a meeting} is associated with meeting effectiveness and inclusiveness. Meetings where everyone participates are rated as more effective and inclusive than those where this isn't the case.

This confirms that the choice of \texttt{Participation} for an attendee not only impacts her own meeting experience but also can influence the collective experience of all attendees in the meeting. This aligns with research suggesting that the choice to participate is influenced by social norms \cite{baym_collaboration_2021}, where observing participation from others may create a safe space for oneself to participate, thereby increasing the perception of inclusiveness, and ultimately effectiveness. Indeed, Constantinides et al.~\cite{constantinides_comfeel_2020} found that perceiving one's peers to be ``comfortable sharing their thoughts and making contributions'' is related to one's comfort to contribute in a meeting; both are aspects of inclusiveness or psychological safety.

\subsection{Generalizability and Predictive Power}\label{sec:gen_predic}
In this section, beyond identifying a set of meeting attributes that are associated with meeting effectiveness and inclusiveness, we address two key questions that each determine the wider value of our approach and findings. Firstly, we test whether the EIM graph model is generalizable across organizations with different teams, practices, and aims, and therefore different meeting experiences. This is crucial because it tests the wider generalizability not only of our meeting-related insights but also of our measurement and modeling approach more broadly. Secondly, we test whether the available telemetry data we consider is sufficient to reliably \textit{predict} the survey responses to the effectiveness and inclusiveness questions for specific attendees. This is important because such predictability would enable avoiding surveys altogether, thereby minimizing any potential observer effect (see also \ref{sec:limitations}) and organizational burden, and enabling a wider deployment of our approach at scale. 

\subsubsection{EIM Graph Generalizability}
The purpose of the EIM graph model is its descriptive capacity to provide a data-driven structure of correlated attributes of \texttt{Effective} and \texttt{Inclusive} meeting experiences. The insights from this model are reliable and applicable only if this structure does not alter dramatically for a new organization (i.e., the extent of model generalizability). To examine this, we split the data into two different subsets (Subset 1 and Subset 2), defined by different participating organizations, such that the two subsets have a comparable sample size and therefore comparable statistical power. We then fit the EIM graph to these two different subsets and compared the results. 

Table \ref{tab:graphs} shows the three model parameters side-by-side. The parameters of our main graph (Figure \ref{fig:gm}) are under the ``Combined'' column. The parameters of graphs fitted to the two subsets are listed in columns ``Subset 1'' and ``Subset 2''. These graphs have a few edges less than the Combined graph. This makes the Combined model a meta-graph of these two subsets.  This is the result of more edges being pruned (not passing the statistical significance test) during the graph modeling on the two subsets. Other than that, the algorithm suggests exactly the same structure in all three cases: There is no neighboring node of \texttt{Participation}, for example, that would move to the \texttt{Effective} or \texttt{Inclusive} neighborhood. 
Therefore there is no sign indicating that the graph model and insights derived from it cannot be generalized.

The absence of some edges (correlations) in the two subsets is most likely due to the lack of statistical power in sub-samples. This is also demonstrated by the increase in p-values for common parameters when we move from the subset graphs to the Combined model. This pattern emphasizes the importance of sample size and the danger of relying on a small sample for detecting the drivers of meeting experiences. 


\begin{table}
\begin{tabular}{lllll}
\hline
        Target &                  Input &   Combined &    Subset 1 & Subset 2 \\
\hline
     \texttt{Effective} &          \texttt{Quality Issues} &   0.14 (<0.01) &      0.1 (<0.01) &      0.2 (<0.01) \\
     \texttt{Effective} &              \texttt{Inclusive} &  45.48 (<0.01) &       54 (<0.01) &     39.7 (<0.01) \\
     \texttt{Effective} &         \texttt{ScreenShare} &   1.39 (<0.01) &   1.3 (0.05) &  1.39 (0.01) \\
     \texttt{Effective} &   \texttt{Small Meeting (8 or less)} &   1.29 (<0.01) &          - &     1.43 (<0.01) \\
\hline
\texttt{Inclusive} &          \texttt{Quality Issues} &   0.35 (<0.01) &          - &     0.25 (<0.01) \\
     \texttt{Inclusive} &      \texttt{Reliability Issues} &   0.49 (<0.01) &     0.44 (<0.01) &     0.47 (<0.01) \\
     \texttt{Inclusive} &         \texttt{Participation} &   4.05 (<0.01) &     4.47 (<0.01) &     3.58 (<0.01) \\
     \texttt{Inclusive} &   \texttt{Small Meeting (8 or less)} &   1.51 (<0.01) &  1.43 (0.02) &     1.56 (<0.01) \\
     \texttt{Inclusive} &  \texttt{Short Call (10min. or less)} &   0.61 (<0.01) &      0.5 (<0.01) &          - \\
 \hline
 \texttt{Participation} &      \texttt{Reliability Issues} &   0.13 (<0.01) &      0.1 <0.01) &     0.16 (<0.01) \\
 \texttt{Participation} &              \texttt{Recurring} &   0.82 (<0.01) &     0.84 (<0.01) &      0.8 (<0.01) \\
 \texttt{Participation} &         \texttt{ScreenShare} &   0.71 (<0.01) &     0.72 (<0.01) &     0.71 (<0.01) \\
 \texttt{Participation} &   \texttt{Small Meeting (8 Or Less)} &   7.13 (<0.01) &     7.39 (<0.01) &     7.03 (<0.01) \\
 \texttt{Participation} &  \texttt{Short Call (10min. or less} &   0.72 (<0.01) &     0.66 (<0.01) &  0.79 (0.03) \\
 \texttt{Participation} &        \texttt{Headset} &   1.16 (<0.01) &  1.16 (0.01) &  1.16 (0.01) \\
 \texttt{Participation} &    \texttt{Video Duration > 30\%} &   1.17 (<0.01) &  1.17 (0.01) &          - \\
\hline
\end{tabular}
\caption{\label{tab:graphs} Graph parameters are adjusted ORs computed from GLM coefficients. We show the p-values in parentheses next to each parameter. The empty cells (-) show that there is no edge between the corresponding two nodes in that graph. Subset 1 and Subset 2 are mutually exclusive subsets of data in a way that they include non-overlapping participating organizations while maintaining a similar sample size.}
\end{table}

\subsubsection{Predictive Power for EIM Metrics} Our second question concerns the power of the available telemetry data to predict the survey ratings of meeting effectiveness and inclusiveness. Our ultimate aim with predictive modeling is to obviate the need for surveys altogether, so as to reduce potential observer effects and organizational burdens and enable wider deployment of our measurement and modeling approach. Predictive modeling has a different goal and requirements than descriptive modeling. Because our aim is to obviate the need for surveys, our predictive model should not include any survey-based feature data. Moreover, not being bound by descriptive purposes, there is more flexibility about the type of the model and its complexity when choosing a predictive model (e.g., no constraints on structure). We, therefore, trained two separate models to predict \texttt{Effective} and \texttt{Inclusive} ratings without \texttt{Inclusive} as an input variable. 

In this study, the prediction task is a binary classification where a lightGBM model predicts the probability of an \texttt{Effective} (or \texttt{Inclusive}) rating. All available telemetry and some engineered features from telemetry are predictors (i.e., independent variables). These add up to 40 different inputs for each model. Table \ref{tab:pred_var_list} provides the details about these features.

To measure the predictive power of each model, we use the Area Under the Curve (AUC) of the Receiver Operating Characteristic curve. We follow a random sampling cross-validation strategy where we repeatedly split the data into random train and test subsets, and record the model AUC on the test set. Table \ref{tab:auc_pred_power} contains the cross-validation result of 50 random samplings. The results show that there is more power in predicting \texttt{Inclusive} than \texttt{Effective} ratings. However, neither of the models demonstrates a high enough AUC for reliably predicting individual ratings in the absence of a subjective survey. We should note that the unit of random splits is single ratings and not any unique identifier of users. Therefore, it is possible that ratings from the same person on two different meetings fall in train and test, which is a potential data leak that can cause an overestimation of AUC, i.e., the current AUC is the best value that this data can provide. We can still safely conclude that meeting telemetry by itself is not adequate to predict individual ratings. 

\begin{table}
    \centering
    \begin{tabular}{lc}
    \hline
         & AUC +/- error \\
         \hline\hline
         \texttt{Effective} & 0.65+/-0.02 \\
         \texttt{Inclusive} & 0.72+/-0.02 \\
         \hline
    \end{tabular}
    \caption{Cross-validation AUC on 50 randomly selected test sets for models predicting \texttt{Inclusive} than \texttt{Effective} ratings using available telemetry only.}
    \label{tab:auc_pred_power}
\end{table}

Next, we measure the change in AUC when the model moves to predict ratings for meetings outside the organizations in the training set. This is especially critical for being able to use the model in a widely deployed CMC system to automatically predict ratings for meetings without survey ratings. To test the model performance in this scenario, we run the cross-validation while we keep an entire organization outside the training and test data. We repeat this process with three different organizations. The results show a small but statistically significant drop in AUC for both \texttt{Effective} and \texttt{Inclusive} models (see Table \ref{tab:auc_gnr}). That means the models' accuracy in predicting survey ratings will reduce slightly if the model is used to predict ratings in an organization that is not part of the training set. Note that the average AUC on training cross-validation in Table \ref{tab:auc_gnr} is lower than in Table \ref{tab:auc_pred_power} because the training dataset in the former analysis of generalizability is smaller and less diverse.

Predicting individual effectiveness and inclusiveness ratings is statistically more challenging than deriving common patterns of correlations. The former requires more data points and a broader set of predictive variables. Our study shows that meeting telemetry can provide a robust understanding of the common factors related to effectiveness and inclusiveness. However, without further research, telemetry alone cannot replace a subjective survey by predicting individual ratings. This emphasizes the importance and, by extension, the quality of the survey-based measurement of meeting effectiveness and inclusiveness; Section \ref{sec:skew} below details our work to understand and improve the survey data. 

\begin{table}[h]
    \centering
    \begin{tabular}{l|c|c|c|c}
    \hline
          \multirow{2}{*}{Unseen Test} & 
          \multicolumn{2}{c}{Cross-validation AUC when training} & 
          \multicolumn{2}{c}{Unseen test set AUC}  \\
         \cline{2-5}
          & \texttt{Inclusive} & \texttt{Effective} & \texttt{Inclusive} & \texttt{Effective} \\
         \hline \hline
         Test organization 1 & 0.68+/-0.01 & 0.60/-0.01 & 0.66 & 0.59 \\
         \hline
         Test organization 2 & 0.70+/-0.01 & 0.63/-0.01 & 0.66 & 0.58 \\
         \hline
         Test organization 3 & 0.69+/-0.01 & 0.62/-0.01 & 0.67 & 0.60 \\
         \hline\hline
         Average & 0.69 & 0.62 & 0.67 & 0.59 \\
         \hline
    \end{tabular}
    \caption{AUC when moving the model to an unseen test set.}
    \label{tab:auc_gnr}
\end{table}

\section{Survey measurement of meeting effectiveness and inclusiveness: Survey skew and its solutions}\label{sec:skew}

The survey-based measurement of meeting effectiveness and inclusiveness is core to our methodology as, to our knowledge, it provides the only tractable way of measuring these subjective and complex constructs (e.g., \cite{constantinides_comfeel_2020}). The survey data, therefore, serves as the ``ground truth'' in our descriptive and predictive modeling, with the value of telemetry attributes depending heavily on the quality of the survey data. Data quality becomes even more important as surveys are deployed in the real-world organizational context. Ensuring high-quality survey data is therefore a key focus. One of the main challenges we observed is participants' tendency to provide 4 or 5-star ratings regardless of their true experience (i.e., what is termed ``skew'' in the data). Table \ref{tab:rating-dist} shows the relative frequencies of star ratings provided by survey respondents; a significant portion of surveys is rated with 4- or 5-star ratings. This section describes studies we conducted to understand the skew in ratings and approaches to mitigate its impact.

This skew pattern has been reported in other settings that rely on user ratings, such as online marketplaces like eBay, Uber, and AirBnB \cite{tadelis_reputation_2016,garg_designing_2019}. There are several potential complementary explanations for this. First, low ratings could incur a perceived reputation cost for yourself and others: low ratings could encourage retaliatory behavior from those being rated or could incur other social costs, particularly in a workplace context \cite{tadelis_reputation_2016}. Raters may therefore avoid low ratings out of fear of this cost to themselves or others. Sections \ref{sec:wordedlabels} and \ref{sec:useranonymity} describe analyses addressing this. Second, due to the ubiquity of star rating systems and aforementioned reputation costs, participants may have learned norms that responding with 4 or 5 stars is the appropriate response regardless of context (see Section \ref{sec:wordedlabels} for more on this) \cite{garg_designing_2019}. Third, and related to the previous reasons, participants may have limited capacity to respond to the survey, so defaulting to the norm of 4 or 5 stars is the fastest and safest approach in terms of reputation and cognitive costs of providing informative ratings (see Sections \ref{sec:completiontime}, \ref{sec:surveyfatigue}, and \ref{sec:meetingtiming} for more on this).

\begin{table}
\begin{tabular}{lcc}
        \hline
        & \texttt{Effective} & \texttt{Inclusive} \\
        \hline
        1-star	& 1\% & 1\%\\
        2-star	& 1\% & 1\%\\
        3-star & 3\% & 3\% \\
        4-star	& 12\% & 10\% \\
        5-star	& 82\% & 86\% \\
        \hline
\end{tabular}
\caption{\label{tab:rating-dist}Relative frequency of star ratings provided by respondents shows a high tendency to rate 4 or 5 stars for meeting \texttt{Effectiveness} and \texttt{Inclusiveness}.}
\end{table}

\subsection{Replacing stars with worded labels}\label{sec:wordedlabels}
\begin{figure}[h]
    \centering
    \includegraphics[width=1\linewidth,trim={0mm 30mm 40mm 0mm},clip]{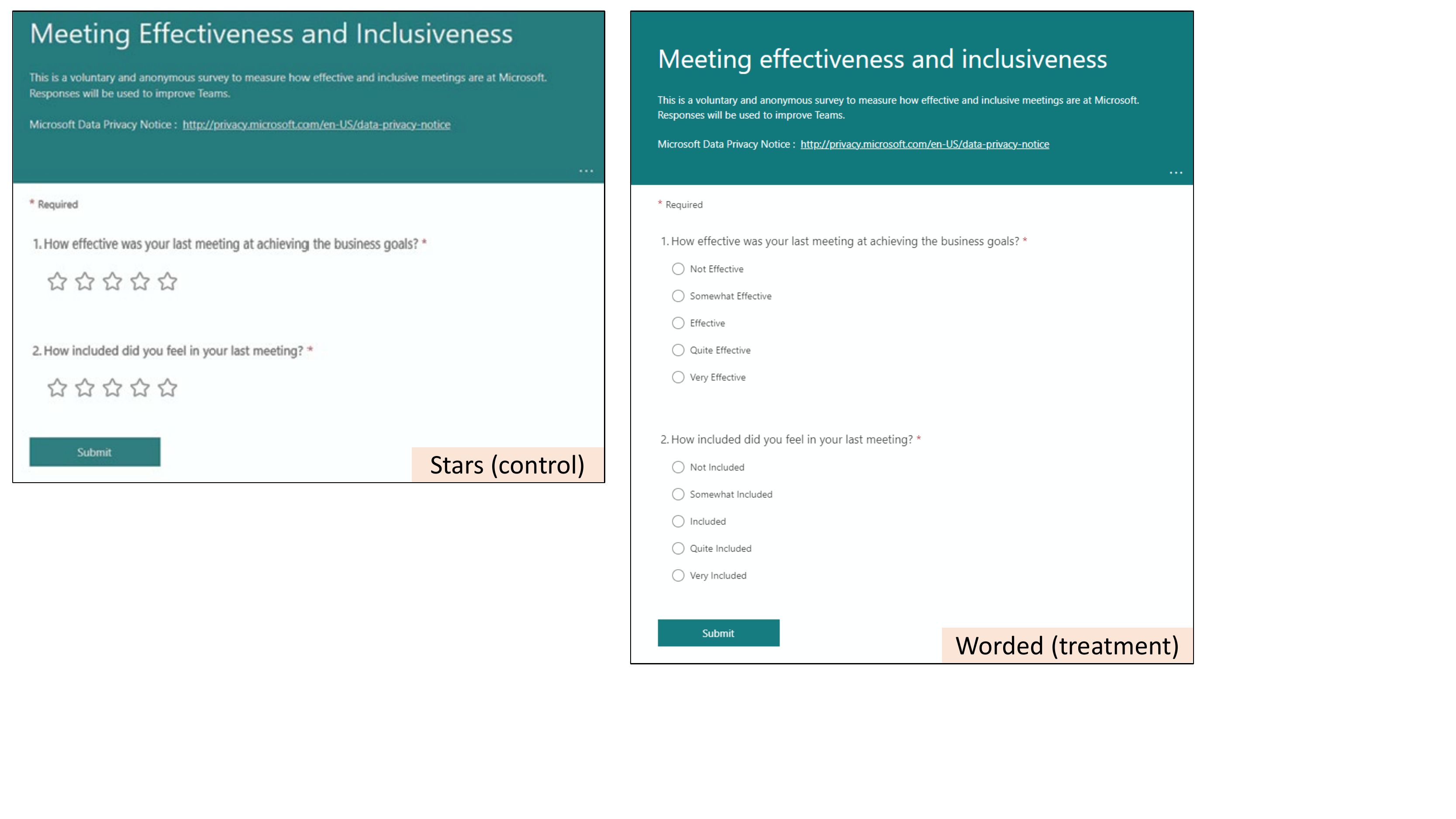}
    \caption{Participants received either a control survey where they rated meeting effectiveness and inclusiveness out of 5 stars or a treatment survey where they rated instead out of positively skewed worded options.}
    \label{fig:skewexp_design}
\end{figure}

One hypothesis is that perceived reputation costs associated with low ratings discourage such ratings (e.g., raters don't want to disparage the meeting organizer), leading participants to provide 4- or 5-star ratings \cite{tadelis_reputation_2016}. This may be further exacerbated by the ubiquity of online star rating systems which reinforce learned norms around how to respond \cite{garg_designing_2019}.  \cite{garg_designing_2019} showed that replacing a star rating system with positive-skewed worded options (response options that have more positive- than negative-valenced verbal labels) substantially reduces rating skew in an online marketplace. First, a positive-skewed option set may reduce the perceived reputation cost (for yourself and others) of providing lower ratings; second, worded options may discourage respondents from relying on learned norms around star ratings. We conducted a small experiment to test this approach in the meeting rating context. A random subset of 2,000 employees was polled via email and randomly allocated to receive a link to one of two surveys as displayed in Figure \ref{fig:skewexp_design} (we relied on email as a rapid approach to initial experimentation). The \textit{control} survey invited the respondents to rate the effectiveness and inclusiveness of their last meeting on a scale of 1 to 5 stars, and the \textit{treatment} survey instead asked respondents to rate the meeting on a scale of ``Not Effective'', ``Somewhat Effective'', ``Effective'', ``Quite Effective'', ``Very Effective'' (with analogous options for inclusiveness). 

\begin{figure}[h]
    \centering
    \subfloat[\centering Effectiveness]{{\includegraphics[width=0.45\textwidth,trim={0mm 0mm 0mm 0mm},clip]{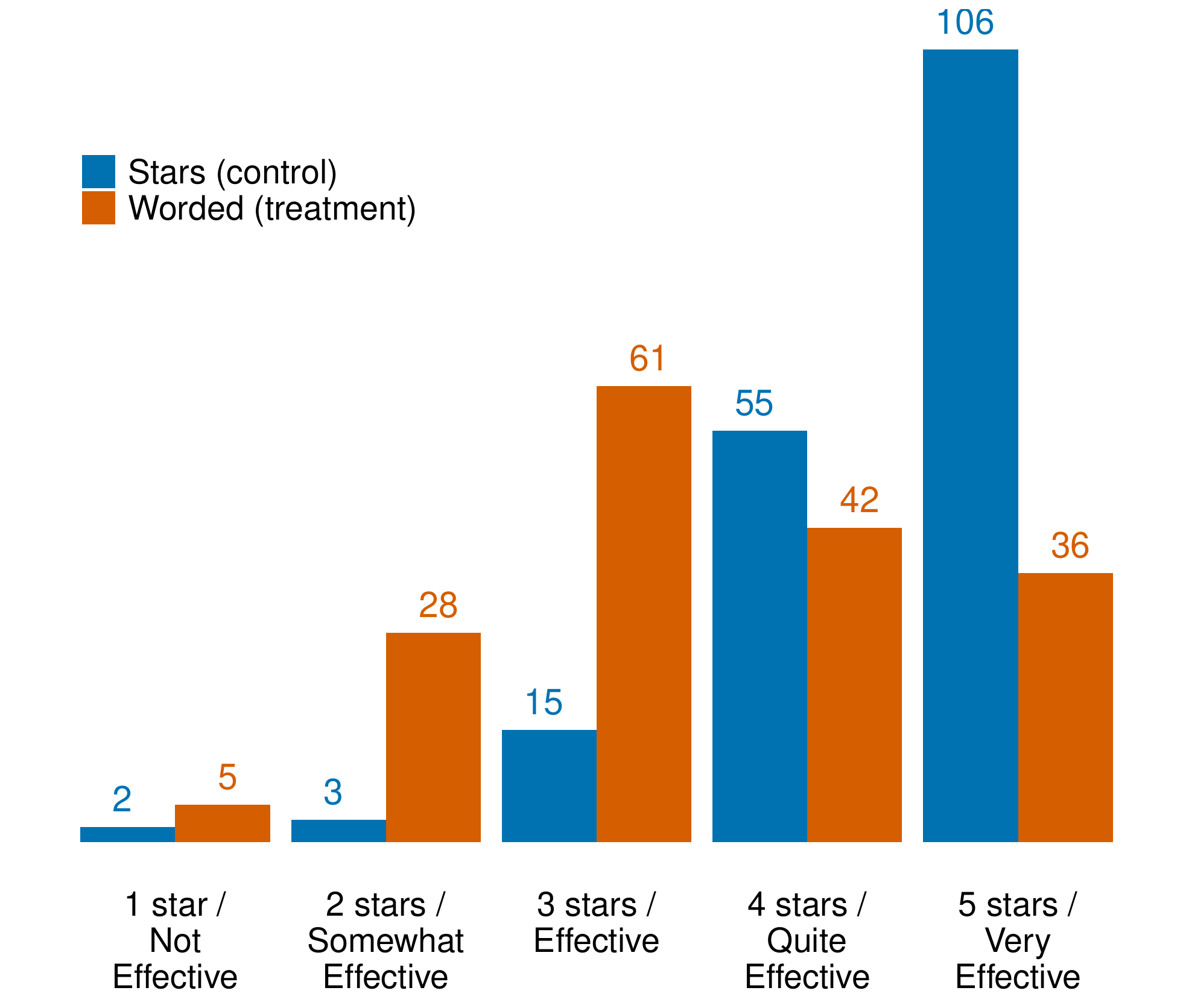} }}
    \qquad
    \subfloat[\centering Inclusiveness]{{\includegraphics[width=0.45\textwidth,trim={0mm 0mm 0mm 0mm},clip]{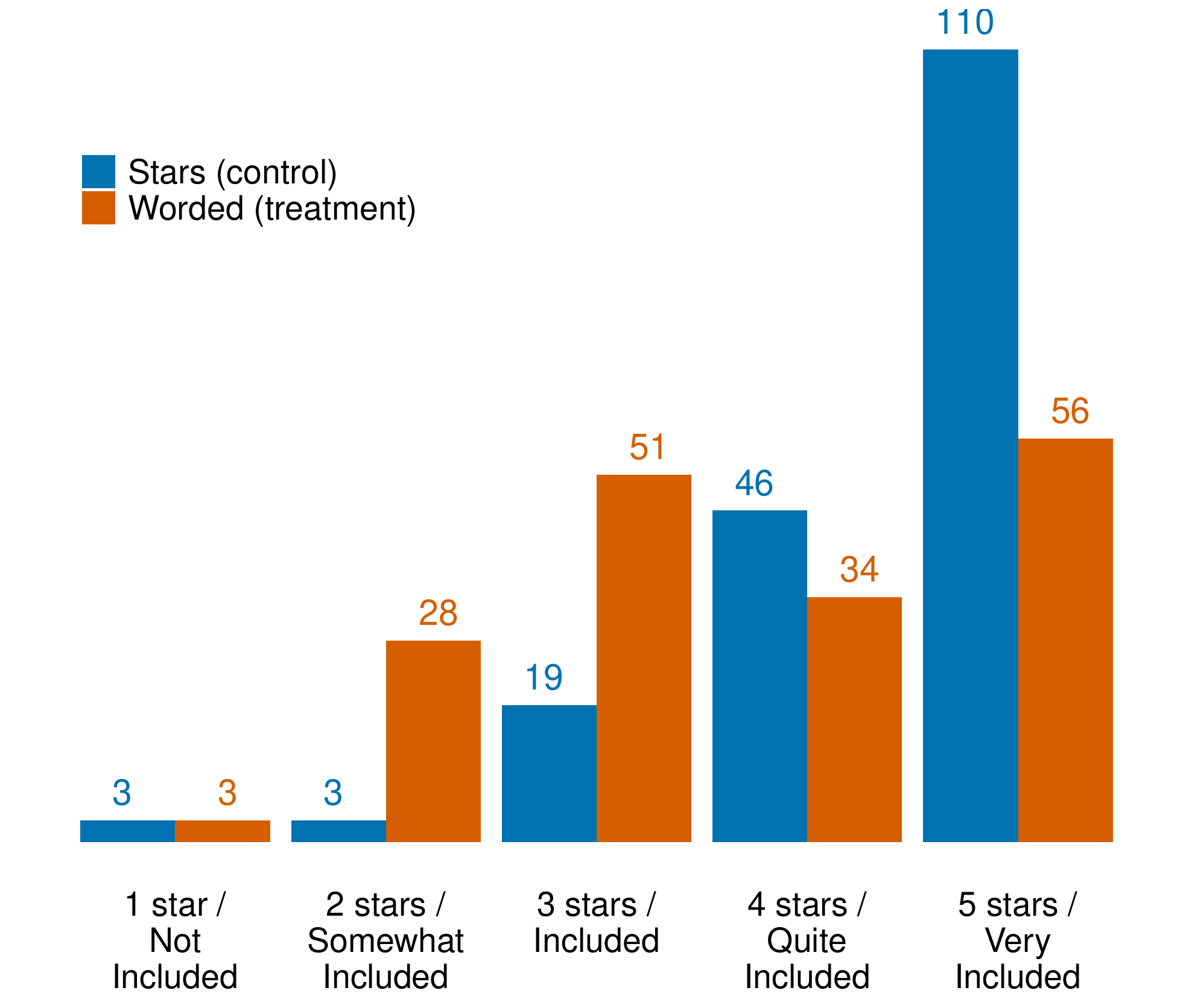} }}
    \caption{Rating skew experiment results for ratings of meeting (a) effectiveness and (b) inclusiveness. Figures show the distribution of responses for the stars (control) condition and the worded (treatment) condition. Numbers indicate the counts of participants selecting that option.}
    \Description{Rating skew experiment results for ratings of meeting (a) effectiveness and (b) inclusiveness. Figures show the distribution of responses for the stars (control) condition and the worded (treatment) condition. Numbers indicate the counts of participants selecting that option.}
    \label{fig:skewexp_results}
\end{figure}

We had a response rate of 18.1\% and 17.2\% for the control and treatment conditions, respectively (no significant difference between these rates using a Fisher's exact test, p = 0.639). When comparing the distribution of responses between conditions, we found a substantial difference for both effectiveness and inclusiveness ratings, as seen in Figure \ref{fig:skewexp_results}. A Fisher's exact test of the counts across ratings showed a statistically significant difference between conditions for both rating questions (p < 0.001 for both). To further estimate the improvement with the worded options, we computed the Shannon Entropy in each condition as a measure of the information in the data. We observed an increase from 1.44 to 2.07 bits for effectiveness (a 44\% increase), and an increase from 1.48 to 2.04 for inclusiveness (a 38\% increase), respectively. Thus, we demonstrate as a proof-of-concept that positive-skewed worded response options can decrease rating skew, relative to star ratings, for ratings of meeting effectiveness and inclusiveness. This is a promising direction and should be further tested in the CMC system when feasible.

\subsection{Survey completion time}\label{sec:completiontime}
Another hypothesis for the cause of the skew was that respondents were responding with 4 or 5 stars too quickly to get through the survey (i.e., they had limited capacity to respond) and those responses weren't reflective of their experience. The work done in \cite{wood_response_2017} examined the appropriateness of using response speed as an indicator of data quality.  We used their recommended filter of 4 seconds for the 2-question survey.   Comparing the skew on the data with the filter of response time, the results showed a reduction of users rating 5  on both inclusive and effectiveness by 14 percentage points absolute.  The rate of 4 or 5 on inclusiveness and effectiveness individually were both reduced by 3\%. Since this filter was able to improve data quality by reducing the skew, it was applied to the data before the analysis and modeling presented in Section \ref{sec:graph}.     

\subsection{Survey fatigue}\label{sec:surveyfatigue}
As the survey was available for a longer period of time we noticed an increase in the rate of 4- and 5-star responses, as well as a decrease in the survey response rate. One suspected reason for these changes 
was user fatigue with the survey. A longitudinal study was conducted by grouping the data based on how long it had been since the user had previously experienced the survey. Then we could look at the rate of responding 5 to both questions for the different groups. There is a bias where users who have more meetings are more likely to observe the survey and have shorter times between survey exposures. 
  To address this bias, we did this analysis comparing people with similar survey exposure counts.  What we found was a critical point where 7 days after receiving the survey there was a decrease in users rating 5 to both questions by 1.6\%.  This supports having a cool-down period between survey requests. This was implemented in our CMC system for this work and executed as detailed in Section \ref{sec:experiment}.

\subsection{Timing of meetings}\label{sec:meetingtiming}
Another scenario potentially contributing to the skew was the timing of the meeting for the user.  We identified for these studies the time of day the meeting occurred if the user had a meeting after the meeting occurred, and if they had multiple meetings that day.  

We did a study on the subset of the data where the time from when the current meeting ended to when the next meeting began was within 2 hours.   On this data, we compared the response rate and metric rates for different times until the next meeting began. We found that when the time to the next meeting is less than 5 minutes from when the questionnaire is shown there is a 2\% lower response rate, and the users were more likely to rate the meeting effectiveness and inclusiveness lower.  This directly shows this is not a cause of the rating skew.  Moreover, there is no need to implement a filter that suppresses the survey in the event of users having meetings close to each other in the CMC system. 

Considering ratings by the time of day that the meeting occurred, we found that there was no statistical difference in the rating distribution.  

The last thing we checked in terms of someone having a busy day was the number of meetings they had using the CMC system that day.  What we found was the more meetings someone had in a day the less likely they were to respond to the survey, and the more likely they were to rate the meeting 5 stars.  We looked for an interaction effect for the time of day for people with large volumes of meetings, but the difference in metric distribution for this group was not significant. 

\subsection{User demographics}\label{sec:userdemographics}
We hypothesized there is a demographic landscape of raters that we are not covering in our modeling.   People with different jobs are expected to have different rating behaviors.  We defined user cohorts based on net user behavior, looking at how the individual engaged with the CMC system over the past month. To do this, we used the total number of meetings, the average meeting size, the frequency at which the user hosts meetings, and the percentage of meetings rated.  There are multiple ways to cluster data. Our priority was to have a clustering that could give us information about different demographics. To maintain information about the clusters we partitioned each feature into 2 bins based on a set of thresholds.   These thresholds are 56 for the number of meetings, 30\% for the percentage of meetings rated, 20\% for the percentage of hosted meetings, and 10 for the average meeting size.  Looking at all possible combinations of these and the resulting metrics there was a clear trend to have 3 cohorts defined as:
\begin{itemize}
    \item Cohort 0: The percentage of rated meetings is low.   
    \item Cohort 1: The percentage of rated meetings is high and the number of meetings is low.
    \item Cohort 2: The percentage of rated meetings is high and the meeting size is large.
\end{itemize}
These cohorts had statistically different distributions of the results. With Cohort 0 showing the least skew, and Cohort 2 showing the most. These cohorts proved to be useful in modeling. However, we were not able to use them for this study given that the cohort computation requires historical data and many of the organizations in our data were relatively new. Future work should explore this further.

\subsection{User anonymity}\label{sec:useranonymity}
User perceptions of meeting inclusiveness and effectiveness are sensitive data to collect. Users might be wary of answering due to the potential impact on themselves or their coworkers.   To test if anonymity was important we conducted an experiment with surveys that were sent via email instead of integrated into the CMC system. Users were sent emails with a link to a survey, we had two surveys with the same questions but with different privacy statements.  The first statement mirrored the survey through the CMC system, providing a link to a data protection notice. The second statement had a stronger worded privacy statement that additionally included that their responses were anonymous.  We found no statistical difference in the rating distribution of the two surveys. This analysis did not provide evidence that letting raters know their responses were anonymous would resolve the skew.

\section{Summary of findings}
In this work, we developed descriptive and predictive models for meeting effectiveness and inclusiveness and showed that they generalize across multiple organizations. We conducted an analysis using the models and found several novel insights about the attributes driving meeting effectiveness and inclusiveness (see Insights in Section \ref{sec:insights}). We also analyzed and shared solutions for the survey skew which is the major data quality challenge in the subjective measurement of meeting effectiveness and inclusiveness. 

We summarize the results of our research in three areas related to the three research questions described in Section \ref{sec:intro}.
\begin{itemize}
    \item CMC systems can implement automated randomized surveys about meeting effectiveness and inclusiveness to show at the end of a meeting. These surveys, deployed at a low triggering rate, can produce a steady flow of subjective measurements. The resulting data is reliable for measuring the overall patterns and changes in meeting experiences over time. The main challenge is data quality, particularly the survey rating skew. We can address rating skew by using worded positively-skewed response options and by implementing a delay time between two consecutive survey exposures for a user. Additionally, considering a minimum response time per question is a powerful approach for post hoc filtering of potentially invalid survey response data. For this purpose, the CMC system should submit the survey completion time alongside other telemetry.
    \item The descriptive model and data confirm that participating in conversations is the most significant factor in meeting inclusiveness and effectiveness, both at the individual and group levels. Beyond that, small meetings are also associated with increased participation. Meetings with \texttt{ScreenShare} are more effective but not necessarily more inclusive. On the other hand, video usage can help with inclusiveness for small and short meetings. 
    \item Statistical models are beneficial for decomposing the subjective ratings of effectiveness and inclusiveness onto telemetry-based attributes. However, their predictive power is limited in predicting individual ratings based on the meeting telemetry data available today alone. The descriptive models are robust and reliable concerning the shifts in the underlying data (e.g., across organizations), but the moderate predictive power of statistical models can suffer significantly when applied to unseen data.
\end{itemize}

\section{Discussion}
\subsection{Theoretical implications}
Understanding and improving meetings has been a central focus of meeting science since the pioneering work of Schwartzman \cite{schwartzman_meeting_1989}. The vast majority of such research has relied on small-scale surveys that often ask about people's general meeting experiences, rather than enabling analysis at the meeting level (e.g., \cite{cohen_meeting_2011,leach_perceived_2009}), or involves small-scale analyses of meetings using manual coding of behavior (e.g., \cite{allen_effective_2015,lehmann-willenbrock_how_2014}). The COVID-19-related shift to increased remote and hybrid work has provided an opportunity to understand and improve meetings at scale using CMC systems. Here we demonstrate the feasibility and value of a large-scale, cross-organizational approach to measuring and linking subjective survey data on meeting effectiveness and inclusiveness, together with objective telemetry data on meeting participation and other attributes captured via a CMC system during real-world remote meetings. 

We posit that large-scale measurement of meetings is key to an in-depth understanding of the factors that contribute to successful meetings. Moreover, it enables organizations to contextually understand their own meetings and rigorously test their own policy-based or technological interventions for improving them, thereby addressing the heterogeneity common to behavioral change \cite{bryan_behavioural_2021}. Indeed, this value is recognized by the organizational executives we partnered with, who wanted a clear line of sight to our model's ability to show statistically significant changes for organizational interventions to improve meeting culture. 

Objective data on meeting participation and other attributes (i.e., telemetry) is central to our approach as it can ultimately enable at-scale and passive measurement of meeting experiences \cite{constantinides_future_2022}. Meeting effectiveness, and especially inclusiveness, are both complex constructs with many subjective elements that are difficult to currently capture using objective measures (e.g., \cite{reinig_toward_2003,niebuhr_sound_2021}); hence, we relied on survey ratings to measure these and conducted in-depth investigations to improve survey data quality. However, the ultimate aim of our approach for real-world deployment is to build a predictive model using objective telemetry that can accurately estimate meeting effectiveness and inclusiveness without the need for survey ratings. To this end, we demonstrate that our descriptive model generalizes well across organizations, but that predicting effectiveness and inclusiveness in the absence of survey ratings requires further work to increase the range of telemetry included in the model, and to improve the quality of survey data (see also Section \ref{sec:limitations}). 

\subsection{Design implications}
\subsubsection{Improving meeting design and technologies}
Our analyses of meeting attributes point to opportunities for improving meeting design and technologies. For example, given the importance of participation for meeting effectiveness and inclusiveness, there is a need to reduce barriers to participation. To this end, recent work has developed a detector for failed speech interruption attempts which can help participants take the floor in conversations \cite{fu_improving_2022}. Similarly, meeting dashboards may also provide participants with insights about their conversations to encourage more equitable participation \cite{samrose_meetingcoach_2021}. To address the negative impact of meeting size, there is an opportunity to nudge meeting organizers during scheduling to help them reflect on their intended participants, their workflow, and meeting goals, and decide whether meeting size can be reduced \cite{mroz_we_2018}. 

Our findings underscore the importance of video use for participation and ultimately inclusiveness in small meetings. Ensuring that all participants can transmit reliable and high-quality video is a top priority for improving CMC systems. Moreover, given the fact that video use is strongly influenced by social norms—people's decision to turn on video is at least partly determined by whether others in the meeting also do so \cite{balogova_how_2022,baym_collaboration_2021,bennett_videoconference_2021}—there is an opportunity to use in-system reminders to encourage video use in relevant meeting contexts. Alternatively, as a mitigation against video fatigue \cite{bennett_videoconference_2021}, there is an opportunity to explore whether avatars can serve a similar purpose in increasing social presence and meeting participation \cite{panda_alltogether_2022, wong_effects_2022} (see also Section \ref{sec:limitations}).  

\subsubsection{Deployment of large-scale meeting measurement systems}
Deploying large-scale measurement and modeling of behavior into organizations invites understandable concerns and requires buy-in. We encountered several challenges in the setup of the study. First, survey respondents needed more clarity about how responses would be used, reflecting common concerns about workplace surveillance (which can include hierarchical as well as peer-to-peer surveillance \cite{soga_web_2021,andrejevic_work_2002}), particularly in the context of the automated collection of telemetry data. If deployed inappropriately and without employee buy-in, workplace surveillance has the potential to reduce beliefs in organizational fairness, trust in leadership, and commitment \cite{chory_organizational_2016}, as well as increase stress \cite{ravid_meta-analysis_2020}. Considering the (explicit and perceived) purpose, invasiveness, frequency and regularity, and transparency of large-scale meeting measurement systems are essential prior to wider deployment \cite{ravid_epm_2020}. Secondly, as mentioned above, organizational executives wanted confidence in the models’ ability to show statistically significant changes associated with changes they made to meeting culture. Relatedly, there were concerns about the model's explainability to external stakeholders. Indeed, explainability is important for justifying decisions (particularly if meeting measurement systems become tied to employee performance metrics), and increasing understanding and therefore control of how a system operates \cite{adadi_peeking_2018}. Lastly, during survey piloting, external organizational stakeholders in Human Resources and IT were not comfortable with sending company-wide surveys via mass emails. Alternatives like posting the survey to internal social channels resulted in a minimal response rate (less than 3\%), preventing our ability to collect a diverse baseline using this methodology. Sufficient lead time to acquire organizational buy-in, wider considerations of employee rights and perceptions, and iterative testing and feedback are therefore important for the successful deployment of such systems.  

\subsection{Limitations and future research}\label{sec:limitations}
Although we have explored multiple important meeting attributes, there remain further opportunities to improve the model's predictive power by including richer telemetry data. First, participation in remote meetings also includes the use of chat and reactions, which are used in ways that can contribute to effectiveness (e.g., by providing easy ways to share relevant information without disrupting ongoing conversations) and inclusiveness (e.g., by widening participation opportunities for those that may not be able or feel comfortable to participate vocally) \cite{sarkar_promise_2021}. 

Second, prior survey research has found that the impact of participation on meeting outcomes depends on the content and context of participation: meeting citizenship behaviors, such as sharing helpful information or ideas, can improve perceived effectiveness and overall engagement, whereas counterproductive behaviors, such as criticizing others or complaining, can harm it \cite{lehmann-willenbrock_our_2016, a._allen_linking_2014,odermatt_incivility_2018}. Those with meeting-relevant knowledge are more likely to participate and thereby perceive their meetings as being more effective, particularly in meeting contexts with high participant disagreement \cite{lindquist_if_2020}. For privacy reasons, our data does not have information on the content of verbal participation, yet recent research shows that such content may be particularly predictive of meeting effectiveness and inclusiveness \cite{zhou_role_2021, zhou_predicting_2022}. Future work should therefore consider how to leverage such data in a privacy-preserving manner (see also below).

Third, the dynamics of participation, including patterns in turn-taking and other aspects of conversation flow, can yield rich insights into effectiveness and inclusiveness in a privacy-preserving manner (i.e., without considering the content of speech) \cite{fu_improving_2022, margariti_automated_2022, reece_advancing_2022}. Potentially fruitful data here includes patterns in the timing and duration of speech across participants \cite{margariti_automated_2022}, choral responses like laughter \cite{bonin_coginfocom_2012}, acoustic features like prosody \cite{niebuhr_sound_2021}, gestures such as head nods \cite{samrose_meetingcoach_2021}, and eye gaze patterns \cite{degutyte_role_2021}. 

Fourth, as mentioned above, avatars and other mixed reality technologies provide alternative media choices for participants to convey their presence in meetings without relying on video \cite{panda_alltogether_2022}. Understanding the impact of these choices on effectiveness and inclusiveness, and the interaction with video use will be important.

Relatedly, hybrid meetings (with both on-site and remote attendees) pose another challenge and opportunity for our methodology which has so far focused on remote attendance (though the majority of meetings in our data are likely to be all-remote, our data cannot identify participants that were co-located in hybrid meetings). The user experience for post-meeting surveys requires careful consideration about where and when to deliver surveys for on-site meeting attendees, ensuring that survey responding is easy and privacy-preserving for everyone. Additionally, telemetry requires further processing to accommodate multiple on-site attendees (e.g., speaker diarization to accurately measure vocal participation for each person). Given the social and interaction asymmetries common to hybrid meetings, measuring meeting effectiveness and inclusiveness for such meetings is a priority \cite{nakanishi_hybrid_2019-1}.  

Fifth, there is an opportunity to integrate telemetry about the presence of agendas or other pre-meeting materials, action points, and other post-meeting minutes that are known to be key for meeting effectiveness \cite{allen_key_2022,mroz_we_2018,constantinides_comfeel_2020}. For example, a lack of clear goals, agendas, and post-meeting summaries have been shown to be negatively correlated with meeting quality \cite{cohen_meeting_2011} and meeting effectiveness \cite{garcia_understanding_2019,leach_perceived_2009, geimer_meetings_2015}.

Lastly, we expect that meeting type (whether a meeting is for brainstorming, decision-making, etc.) has a significant role in modeling meeting effectiveness and inclusiveness \cite{allen_key_2022}. However, telemetry does not contain such information, and current machine learning (ML) solutions cannot reliably extract it automatically from the meeting invite. Making this information available has a high potential for improving EIM model accuracy.  
 
While telemetry provides opportunities for more accurate measurements, it also imposes limitations on the scope of available attributes. There are aspects (such as participant demographics) that are expected to have statistical predictive power but are not available to a CMC system due to privacy and security concerns. This can prevent the model's ability to achieve 100\% descriptive or predictive power of individual subjective assessments of effectiveness and inclusiveness. However, it also enables any ML solution based on this data to be more secure and fair in real-world applications.

As explored in Section \ref{sec:skew}, one major challenge in the subjective measurement of meeting effectiveness and inclusiveness is reducing the commonly observed skew of ratings. Survey design, such as the design of response options and the timing of deployment, is key for mitigating this; data can be refined further by applying appropriate filters (e.g., based on response times). Other opportunities include using nudges based on social norms or incentives to encourage participants to respond more frequently and honestly \cite{chung_social_2016, smith_effectiveness_2019}. Future work should test these ideas in new contexts and at scale, and explore other ways of improving data quality. 

The ``observer effect'' poses an additional potential challenge to our methodology \cite{macefield_usability_2007}. Participants' meeting interactions and survey responses may be influenced by their awareness of being observed and measured via surveys and telemetry. Although we cannot exclude this possibility, several lines of evidence argue against it. Firstly, as discussed in Section \ref{sec:useranonymity}, informing participants that their survey responses would be anonymous, thereby reducing the expectation of them being specifically observed, did not influence their ratings. Secondly, other than during the study introduction and the post-meeting surveys, participants had no indication that meetings were being measured (including via telemetry, which includes only standard data regularly logged for engineering purposes). Thirdly, we chose a relatively low survey frequency (10\% of meetings) and a 7-day cool-down period to minimize the impact of the survey on user behavior.
Moreover, neither participants nor their managers were provided with the results of the survey ratings or telemetry during the study period, thereby precluding the influence of feedback on performance and subsequent ratings. Lastly, our observed skewed distribution of survey ratings is similar to that observed in online platforms where users likely do not have strong expectations of being observed by experimenters (at least during normal use) \cite{tadelis_reputation_2016}. Similarly, given that the study was conducted over four months in a real-world context, participants may have habituated to the study, thereby reducing any potential observer effects. Future work could further estimate the extent of the observer effect in the current context, for example, by testing whether pre-meeting or pre-study information about subsequent measurement influences participant behavior. 

Lastly, our methodology requires participants to voluntarily complete the surveys, which opens up the potential for a self-selection bias in the study sample. That is, the people who took the time to complete the surveys may be systematically different from the overall population of employees (e.g., they may be less busy and/or more engaged in the workplace than the overall population). Indeed, as Section \ref{sec:userdemographics} suggests, there are differences in rating behavior based on factors such as how many meetings a person organizes and the size of their meetings. The privacy-preserving nature of our methodology precluded us from capturing more detailed demographics. However, future work could estimate and mitigate self-selection bias by capturing relevant demographic variables about participants, comparing them to the demographics of the overall population, and targeting surveys accordingly to minimize bias. Ultimately, improving the modeling of meeting effectiveness and inclusiveness using telemetry will decrease the need for self-selected survey data.

\begin{acks}
  We thank Scott Inglis, Thierry Tremblay, and Dejan Ivkovic for their valuable work that enabled survey development and data collection for this reseach.
\end{acks}
  

\bibliographystyle{ACM-Reference-Format}
\bibliography{IC3-AI}

\appendix
\section*{APPENDIX}
\section{GLM Tables for Side Models}
The tables shared in this section show detailed effects and interaction effects of a subset of attributes on effectiveness and inclusiveness. These subsets are suggested by the sub-graphs that emerged from graph modeling fitted on all attributes.

\begin{table}[h]
\begin{tabularx}{0.9\textwidth}{l c c c}
& Coef & Standardized Coef & p-value \\ 
\hline\hline
 Intercept & 3.80 & 37.79 & 0.00 \\ 
 \texttt{Short Call (30min or less)} & -0.27 & -2.39 & 0.02 \\
 \texttt{Meeting Size} & -0.06 & -8.63 & 0.00 \\
 \texttt{Short Call (30min or less)} : \texttt{Meeting Size} & 0.00 & -0.50 & 0.62 \\
 \texttt{Recurring} & -0.37 & -3.24 & 0.00 \\ 
 \texttt{Recurring} : \texttt{Meeting Size} & 0.03 & 3.41 & 0.00 \\
\hline
\end{tabularx}
\caption{GLM results in modeling the probability of \texttt{Effective} by \texttt{Participation}}
\label{tab:glm_par_eff}
\end{table}

\begin{table}[h]
\begin{tabularx}{0.95\textwidth}{l c c c}
& Coef & Standardized Coef & p-value \\  
\hline\hline
Intercept & -0.40 & -6.90 & 0.00 \\ 
\texttt{Meeting Size (8 or less)} & 2.00 & 2.00 & 0.02 \\
\texttt{VideoDuration30\%} & 0.16 & 1.84 & 0.06 \\
\texttt{CallDuration} & 0.00 & 4.22 & 0.00 \\
\texttt{Meeting Size (8 or less)} : \texttt{VideoDuration30\%} & 0.46 & 5.52 & 0.00 \\ 
\texttt{VideoDuration30\%} : \texttt{CallDuration}  & 0.00 & -2.05 & 0.04 \\ 
\texttt{Meeting Size (8 or less)} : \texttt{VideoDuration30\%} & 0.00 & -2.00 & 0.05 \\
\hline 
\end{tabularx}
\caption{GLM results in modeling the probability of \texttt{Participation} by Video usage}
\label{tab:glm_vid_par}
\end{table}

\begin{figure}[h]
 \centering
 \includegraphics[width=0.6\linewidth]{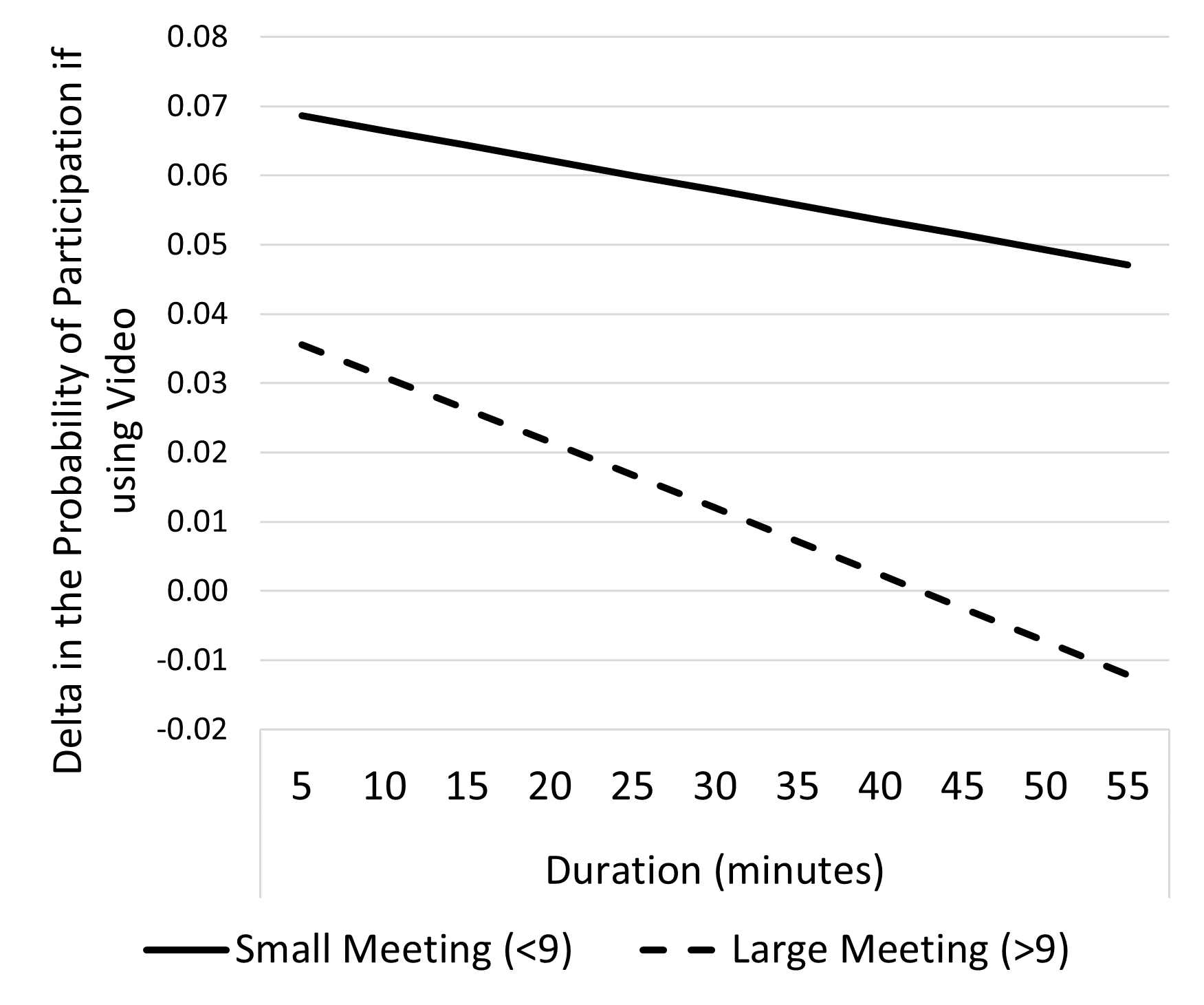}
 \caption{Interaction effect between \texttt{Call Duration}, \texttt{Meeting Size}, \texttt{Video usage}, and \texttt{Participation}.}  
 \label{fig:Participation}
\end{figure}

\begin{table}[h]
\begin{tabularx}{0.9 \textwidth}{l c}
 Name & Type \\  
\hline\hline
\texttt{Microphone Failure (Initialization)}         &     bool\\
    \texttt{Microphone Failure (Mid-Call)}               &     bool\\
    \texttt{Media Failure}                &     bool\\
    \texttt{Media Ever Flowed}         &     bool\\
    \texttt{Reconnect Failure}               &     bool\\
    \texttt{Reconnect Issue}                 &     bool\\
    \texttt{Call Dropped}                      &     bool\\
    \texttt{Video Duration Percent}                &  float\\
    \texttt{Audio Only}                       &     bool\\
    \texttt{Video Only}                        &     bool\\
    \texttt{Video or ScreenShare}                          &     bool\\
    \texttt{ScreenShare Only}                           &     bool\\
    \texttt{Audio Participation Rate} (based on audio packet decoding counts) &  float\\
    \texttt{Is Rater The Meeting Host}                           &     bool\\
    \texttt{Is Friday}                         &     bool\\
    \texttt{is Monday}                         &     bool\\
    \texttt{Is Country GroupA}                        &     bool\\
    \texttt{Is Country GroupB}                        &     bool\\
    \texttt{Meeting Size}                  &    int\\
    \texttt{Call Duration}                      &  float\\
    \texttt{Predicted Probability of Call Quality Issues}                       &  float\\
    \texttt{Total Time In Meeting In The Same Day}             &  float\\
    \texttt{Total Calls In The Same Day}                    &    int\\
    \texttt{ScreenShare > 10\%}             &     bool\\
    \texttt{Quality Issues}                    &    bool\\
    \texttt{Reliability Issues}                &    bool\\
    \texttt{Participation}                   &    int\\
    \texttt{Video Duration > 30\%}               &     bool\\
    \texttt{Recurring}                     &    bool\\
    \texttt{ScreenShare}                  &    bool \\
    \texttt{Small Meeting (8 or less)}             &    bool \\
    \texttt{Short Call (10min. or less)}           &    bool \\
    \texttt{Long Call (1hr or more)}               &    bool \\
    \texttt{Headset}             &    bool \\
    \texttt{Busy Day (10 or More Calls)}            &     bool \\
    \texttt{Short Hours in Meetings (Less Than 1hr In Calls On The Same Day)} & bool \\
\hline 
\end{tabularx}
\caption{List of variables used for predictive modeling.}
\label{tab:pred_var_list}
\end{table}

\received{January 2023}
\received[revised]{July 2023}
\received[accepted]{November 2023}
\end{document}